\documentclass[authoryear,preprint,review,2.5p,12pt]{elsarticle} 
\usepackage[a4paper, margin=1in, left=1in, right=1in]{geometry}

\usepackage{amsmath}
\usepackage{moreverb,url}

\usepackage[utf8]{inputenc}
\usepackage{import}
\usepackage{siunitx}
\usepackage{amsmath,amsfonts,amssymb}
\usepackage{bookmark}
\usepackage{lscape}
\usepackage{graphicx}
\usepackage{setspace}
\usepackage{caption}
\usepackage{subcaption}
\usepackage{tabto}
\usepackage{algpseudocode}
\usepackage{algorithm}
\usepackage{booktabs}
\usepackage{soul}
\usepackage[T1]{fontenc}
\usepackage{subcaption}
\usepackage{hyperref}
\usepackage{tikz}
\setcitestyle{square,numbers,sort&compress,comma}
\newcommand{\rev}[1]{\textcolor{black}{#1}} 

\graphicspath{{Figures/}}

\makeatletter
\def\input@path{{Figures/}}
\makeatother

\usepackage{ulem}
\newcommand{\rn}[1]{\textcolor{black}{#1}} 
\newcommand{\rx}[1]{}

\newcommand\BibTeX{{\rmfamily B\kern-.05em \textsc{i\kern-.025em b}\kern-.08em
T\kern-.1667em\lower.7ex\hbox{E}\kern-.125emX}}

\begin{document}
\begin{frontmatter}

\title{Acoustic Black Hole Damper for Thermoacoustic Instability Control in a Hydrogen Combustor}

\author{Bayu Dharmaputra\corref{cor1}}
\ead{bayud@ethz.ch}
\author{Klejsi Curumi\corref{}}

\author{Nicolas Noiray\corref{cor1}}
\ead{noirayn@ethz.ch}
\cortext[cor1]{Corresponding authors}

\address{CAPS Laboratory, Department of Mechanical and Process Engineering, ETH Z\unexpanded{\"u}rich, 8092, Z\unexpanded{\"u}rich, Switzerland}

\begin{abstract}

Thermoacoustic instabilities remain a major challenge in the operation and development of modern gas turbine combustors for power generation and propulsion. In laboratory environments, such instabilities can also hinder the accurate characterization of key \rx{flame–acoustic}\rn{thermoacoustic} properties \rn{ of the flames}, such as \rx{flame}\rn{their acoustic and} and entropy transfer functions\rn{, because the latter quantities} cannot be reliably measured under unstable conditions. While a variety of active and passive control strategies have been investigated \rn{in the past decades}, passive approaches are generally preferred in industrial gas turbines. Many modern combustors therefore employ wall-mounted acoustic \rx{resonators}\rn{dampers}, such as Helmholtz or quarter-wave resonators; however, these devices are typically effective only over narrow frequency ranges. In this study, the application of perforated acoustic black holes (ABHs) as broadband passive dampers for thermoacoustic instability mitigation is investigated. Several ABH designs are additively manufactured and experimentally characterized through scattering matrix measurements. A reduced-order model based on the transfer matrix method (TMM) is developed and is shown to be in good agreement with the experimental results. Using this validated model, a damper design is optimized to maximize acoustic dissipation over the frequency range 500–2000~Hz, such that the unstable mode of a laboratory-scale technically-premixed hydrogen combustor falls within the effective bandwidth. The optimized perforated ABH damper is installed in the plenum section of the combustor test rig, and its thermoacoustic performance is evaluated over a range of equivalence ratios and outlet boundary conditions. Across all operating conditions considered, the ABH damper leads to a substantial reduction \rx{in}\rn{of the amplitude of the} acoustic pressure oscillation\rn{s}\rx{ amplitudes}. These results demonstrate the potential of perforated ABH-based dampers as a robust and broadband passive solution for mitigating thermoacoustic instabilities in hydrogen-fueled combustors.

\end{abstract}

\end{frontmatter}

\section{Introduction}
With the increasing adoption of hydrogen as an energy carrier to \rn{help} reduc\rx{e}\rn{ing} $\mathrm{CO_2}$ emissions, modern gas turbines\rn{, acting as dispatchable sources for compensating the intermittency of renewables ones,} are increasingly expected to operate with high levels of hydrogen blending \cite{Noble2021,OBERG2022,Ciani2021HydrogenBlending,Aoki2024HydrogenMicromix}. In parallel, future medium-range aircraft are also anticipated to rely on hydrogen-based propulsion systems to decarbonize the aviation sector \cite{FAUREBEAULIEU2024,Clemen2024HydrogenAeroGT}. A critical challenge associated with these developments is the increased susceptibility of hydrogen-fueled combustors to thermoacoustic instabilities, which can arise both during the development phase and under operational conditions. These instabilities result from constructive coupling between unsteady flame heat release and the acoustic field of the combustor, leading to large-amplitude pressure \rn{acoustic} oscillations that can compromise engine integrity \rn{due to the vibration they induce}. Compared to conventional hydrocarbon fuels, hydrogen-blended flames are typically shorter, resulting in significantly reduced characteristic time delays of the thermoacoustic source term\rn{, e.g.} \cite{Moon2024,DHARMAPUTRA2025}. Moreover, owing to the high reactivity of hydrogen, such flames exhibit higher extinction strain rates and are therefore more resilient to velocity perturbations. Consequently, the heat release rate tends to saturate only at higher perturbation amplitudes than in less reactive fuels \cite{Eirik2023}, which can promote the development of larger \rn{acoustic} pressure oscillation amplitudes, as observed experimentally in \cite{AGUILAR2022}. In addition, the reduced flame length associated with hydrogen combustion leads to higher cut-off frequencies of the flame acoustic response \cite{Eirik2023}. Recent studies have shown that hydrogen flames anchored on aeroengine burners can exhibit non-negligible acoustic gain at frequencies extending up to 2000~Hz \cite{FAUREBEAULIEU2024}. As a result, variations in fuel composition further complicate combustor dynamics, since changes in hydrogen content directly affect flame length, time delay, and the frequency range over which flame-acoustic coupling remains significant. In particular, increasing the hydrogen fraction shortens the flame, decreases the effective time delay, raises the cut-off frequency of the flame response, and can therefore shift thermoacoustic instabilities toward higher frequencies \cite{LEE2020}. Achieving fuel flexibility in gas turbines thus requires robust control strategies capable of stabilizing the combustor over a broad range of operating conditions and instability frequencies. In academic settings, thermoacoustic instabilities can hinder the proper characterization of key operating conditions; for instance, quantities such as the entropy and flame transfer functions cannot be measured if the system is \rn{thermoacoustically} unstable.

 \rx{Active and }Passive \rn{and active} control for thermoacoustic instabilities mitigation have been studied in the literature\rn{, e.g. }\cite{Bourquard2019288,DHARMAPUTRA2024}. \rx{T}\rn{So far, t}he need of high-speed and robust \rn{mechanical} actuators \rn{has} hinder\rn{ed} the \rx{real}\rn{practical} implementation of active control \rn{strategies}. Recent studies on plasma-based actuation via Nanosecond Repetitively Pulsed Discharges (NRPDs) \cite{DHARMAPUTRA2024_PROCI,DHARMAPUTRA2024,ARAVIND2024} have shown promising results for \rn{such electric actuator based} active control \rn{system}. However, further research is still needed to \rx{assess}\rn{ensure} the long-term durability of \rx{such an actuator for real implementation}\rn{the associated electrodes}. In principle, both active and passive control \rx{must}\rn{can} coexist and complement each other. \rn{So far, p}\rx{P}assive control methods have been \rx{dominating}\rn{the best practical solutions for}\rx{in the} real gas turbine combustors due to their simplicity. Increasing acoustic losses is one of the most popular passive control strategies, in which quarter-wave and Helmholtz resonators are typically employed. Pandalai \textit{et al.} \cite{Pandalai1998} showed the implementation of a quarter wave resonator mounted in the plenum section of a GE aeroderivative engine that logged more than 100\,000 hours of engine operations. In \cite{Bellucci2004}, Bellucci \textit{et al.} introduced a Helmholtz damper\rx{—}\rn{, which was} later implemented in the ALSTOM GT11N2 silo combustor\rx{—}\rn{, }and developed a nonlinear model to \rx{estimate}\rn{predict} the resonator’s loss coefficient and eigenfrequency. In their study, the target frequency was about 200~Hz, and \rn{the frequency band over which the} \rx{effective} attenuation was \rx{achieved} \rn{effective was} only within approximately 5\% of this value (i.e., $\pm 10~\mathrm{Hz}$). \rn{With such narrow-band solution, it is essential to very carefully tune the dampers to the instability frequency before their implementation}. Further optimization of Helmholtz-resonator-based dampers was presented in \cite{Bothien2013}. There, the authors used interconnected cavities to broaden the effective bandwidth and reported validation test results for their subwavelength damper concept in a heavy-duty gas turbine. However, \rx{the study mainly targeted two frequencies below 400~Hz}\rn{they showed that the trade-off between  broadbandness and damping effectiveness was best achieved for pairs of target frequencies that were not too far appart}. Therefore, developing passive damping solutions that deliver genuinely broadband absorption \rx{at higher}\rn{above a threshold} frequenc\rx{ies—}\rn{y, }while remaining compact, robust, and engine-compatible\rx{—}\rn{, }will be increasingly critical for hydrogen-blended gas turbines, \rn{because they are prone to both low and high frequencies thermoacoustic instabilities. This challenge}\rx{which} directly motivates the present study.
 
 Achieving broadband acoustic absorption and minimal reflection remains a major challenge in noise control \cite{Bravo2023}. Recent research has explored more advanced ideas, including retarding structures for waveguide acoustics, commonly termed acoustic black hole (ABH), where \rn{the `effective'} wave speeds gradually decrease\rn{s} \rx{to the point of vanishing}\rn{and ultimately vanishes} \cite{Li2024,Mi2021}. Under such conditions, an incident wave-above a certain frequency- is effectively `trapped', yielding near perfect absorption without relying solely on material damping \cite{Mi2021,Guash}. The ABH concept originated in the vibration control of beams and plates, where a beam is terminated by a \rx{wadge}\rn{wedge} having a power-law \rn{for its} decreasing thickness\rx{ power law}, which slows and prevents the reflection of flexural waves \cite{PELAT2020115316,Krylov2007_ABH,Mironov1988}. 
 
In the context of duct acoustics, acoustic black holes are most commonly implemented as power-law graded terminations, in which a ribbed or corrugated structure protrudes into the duct and its height increases gradually toward the end of the waveguide \cite{GUASCH2017,HRUSKA2024,Mi2021,Petrover2024,ElOuahabi}. This can effectively be modeled as a duct with gradually decreasing area and varying side wall impedance. As Mironov \cite{Mironov} first showed this “one‐dimensional retarding structure” can guide the acoustic perturbations toward a region of infinitesimal \rn{effective} wave speed, thus eliminating reflections at the duct termination. Such ABH ducts have demonstrated excellent performance in reducing reflection over some frequencies without requiring conventional absorbing layers at the termination. Understanding how geometry and impedance grading interact is crucial for designing an effective ABH. Analytical and numerical models exist to predict the pressure and velocity fields, with the Transfer Matrix Method (TMM) being especially convenient for piecewise segmented ducts \cite{GUASCH2017,Bravo2023}. By discretizing an ABH into a series of cells with gradually varying side wall impedance, the TMM captures both reflection and transmission at each interface and the overall absorption or dissipation characteristics. The TMM method is also well known in the thermoacoustics community as it can serve as a low-order model to predict the thermoacoustic instabilities of a given system \cite{Schuermans2004,Bellucci2005}. 

Conventional waveguide ABH is not compatible with ducted flow applications as there should be no hard termination, and protruding structures will create significant pressure head loss due to flow impingement. Furthermore, some applications require the cross-section area of the duct to remain constant, which is the case in the present study. Studies by Bravo et. al \cite{Bravo2023} tackled this issue\rx{, in that}\rn{. In their} study, the ABH is mounted on the wall of a waveguide. The ABH consists of gradually increasing side branch quarter-wave resonator whose height follows a certain power-law profile along the longitudinal direction and the diameter of the ABH duct is the same as the tube diameter upstream and downstream of it. In that study, the effective \rx{dissipation}\rn{anechoic} cut-on frequency and the bandwidth of the \rn{ABH} are related to the quarter-wave resonance of the cavities. The authors also showed the effect of mean-flow on the dissipation performance of the device. Later work by the same authors \cite{Bravo2024} addresses improving the device’s flow compliance and extending its low-frequency dissipation bandwidth by incorporating a converging duct section together with side-branch coiled cavities of increasing height. In this study, the authors also rename the concept, shifting from “acoustic black hole” to “rainbow trapping silencers\rn{”}. Maury et.al \cite{Maury2025Perforated} implements a micro-perforated panel on the ABH silencer of the study in \cite{Bravo2023} and named the device ABH with perforated \rx{coating}\rn{linings}. By adding the \rx{\rx{perforator}\rn{perforated plate}s}\rn{linings}, the effective bandwidth is shifted to lower frequencies compared to that in the case without \rn{them}\rx{coating}. The authors noted that the \rx{perforator}\rn{perforated plate} increases the robustness in presence of a low-speed flow. The authors further explain and optimize the device in the subsequent study \cite{Maury2025}.

Regarding terminology, a recent review \cite{PELAT2020115316} notes that “acoustic black hole” is now commonly used to denote power-law-type absorbing terminations or indentations, independent of whether attenuation is purely geometric or assisted by local \rn{dissipative} resonances. The device considered here belongs to the resonance-assisted class and can be interpreted as a wall-mounted, power-law graded acoustic indentation. Although local cavity resonances contribute to the attenuation, the defining feature is the deliberate spatial grading, which induces a progressive and distributed modification of the effective acoustic impedance along the propagation direction. Related graded-resonator concepts are also discussed in the literature under the label “rainbow trapping absorbers” \cite{Jimenez2017}, emphasizing frequency-dependent spatial localization of acoustic energy.


Building on these developments, we design a flow-compliant perforated ABH damper, develop a transfer-matrix-method (TMM) model to predict its acoustic response, and integrate the device into a laboratory-scale combustor with a technically premixed hydrogen burner to assess its effectiveness in mitigating thermoacoustic instabilities. The dampers are additively manufactured from polymer via 3D printing and are therefore installed in the cold section of the rig. The results provide a proof of concept that the proposed damper concept can stabilize the system and reduce high-frequency \rx{pressure}\rn{thermoacoustic self-}oscillations of technically premixed hydrogen flames. To the authors’ knowledge, this is the first time a flow-compliant, perforated ABH (rainbow-trapping-type) damper is integrated into a combustor and evaluated for thermoacoustic instability control.

\section{Background Theory}
\subsection{Transfer Matrix Method}\label{sec:TMM}
In this work, we adopt the transfer matrix method (TMM) to model the behavior of the perforated ABH damper. This method is particularly suitable for analyzing cascaded or layered systems, as it efficiently couples the acoustic state at the input of each element to that at the outlet of the same element, and then composes these relations throughout the assembly.
\begin{figure}[t!]
    \centering
\makebox[\textwidth][c]{%
        \def\svgwidth{1.2\textwidth}
        \input{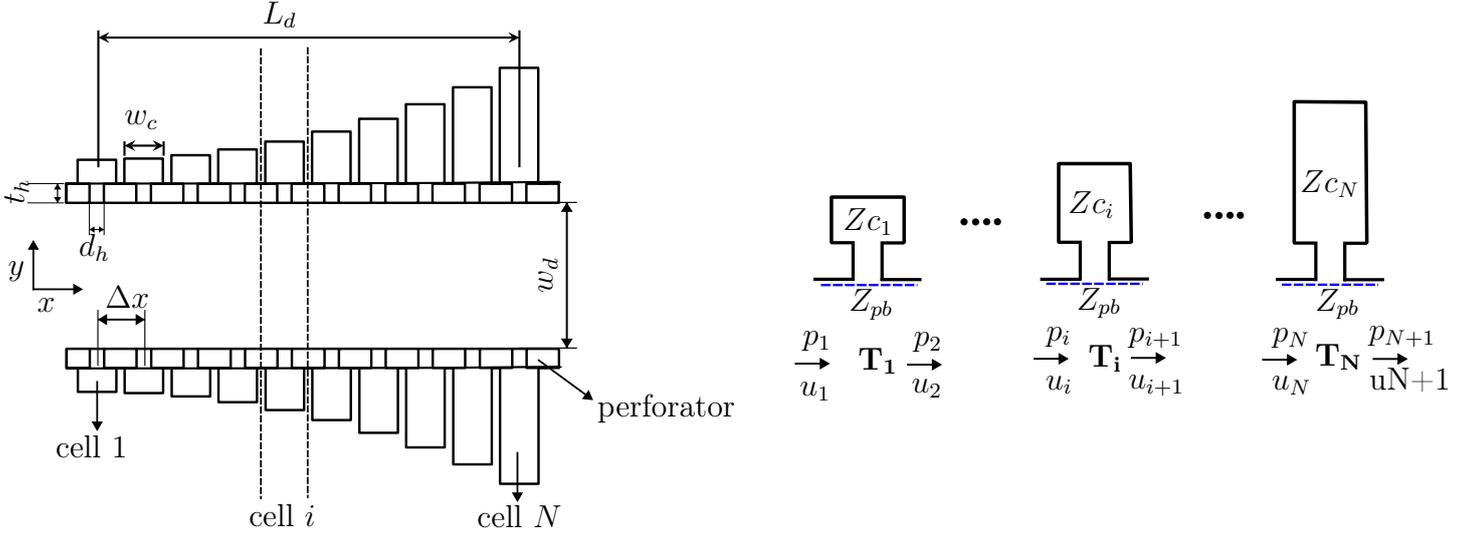}
    }
    \caption{Sketch of the perforated ABH damper including its relevant geometrical parameters (left) and its simplified model for the transfer matrix method (right). }
    \label{fig:TMM}
\end{figure}

Following the methodology of \cite{Bravo2023,Maury2025}, one-dimensional plane-wave propagation is assumed and the structure is discretized into a sequence of elements. Each element consists of a short duct segment, the perforated wall with its backing cavity, and a second short duct segment, as illustrated in Fig.~\ref{fig:TMM}. The full duct region is thus represented by $N$ such elements. Owing to constraints imposed by the experimental combustor setup, the damper is required to have a constant cross-section along the longitudinal direction. The damper cavity height, $H(x)$, increases monotonically along the longitudinal direction according to the following law:

\begin{equation}
    H(x) = H_{min} + (H_{max}-H_{min})\rx{*}\rn{\times}\left(\frac{x}{L_d}\right)^m,\label{eq:profile_func}
\end{equation}
where $H_{min}$ is the minimum cavity height at $x=0$, $H_{max}$ is the cavity height of the last cell, $L_d$ is the center distance between the first and the last holes. The $L_d$ is constrained to 195~mm to make the design compatible with the combustor setup. Each cavity has a width of $w_c$, and the length, $L_c$, is fixed 53.75~mm to make it compatible with the combustor module assembly.

The core idea of the TMM is to relate the acoustic pressure $p$ and the acoustic velocity $u$ at the exit of the $i$-th element, $\bigl[p_{i+1}\;\;u_{i+1}\bigr]^\mathsf{T}$, to those at the inlet, $\bigl[p_i\;\;u_i\bigr]^\mathsf{T}$, by a $2\times2$ transfer matrix $\mathbf{T}_i$:
\begin{equation}
\begin{bmatrix}
p_{i+1} \\[6pt]
u_{i+1}
\end{bmatrix}
=
\mathbf{T}_i
\begin{bmatrix}
p_{i} \\[6pt]
u_{i}
\end{bmatrix}.
\end{equation}
$\mathbf{T}_i$ accounts for the propagation of plane waves within the duct, and possible boundary or impedance conditions such as perforated elements and side cavities. The partial reflections and transmissions at each interface are embedded in $\mathbf{T}_i$ through continuity relations on pressure and velocity. Once the individual $\mathbf{T}_i$ is defined, the global Transfer Matrix $\mathbf{T}_\mathrm{G}$ for the entire discretized structure is simply the product of $\mathbf{T}_i$ from $i=1$ to $N-1$:
\begin{equation}
\mathbf{T}_\mathrm{G} \;=\; \mathbf{T}_{N} \,\mathbf{T}_{N-1} \,\dots\,\mathbf{T}_1.
\label{eq:Global}
\end{equation}
Hence, the state vector at the inlet $[p_1~u_1]^\mathsf{T}$ and the outlet $[p_{N+1}~u_{N+1}]^\mathsf{T}$ of the entire ABH cells are related by
\begin{equation}
\begin{bmatrix}
p_{N+1} \\[3pt]
u_{N+1}
\end{bmatrix}
=
\mathbf{T}_\mathrm{G}
\begin{bmatrix}
p_{1} \\[3pt]
u_{1}
\end{bmatrix}.
\end{equation}
This formulation can then be converted to the scattering matrix, allowing the calculation of the reflection, transmission, and \rx{dissipation}\rn{absorption} characteristics. \rn{Note that in this work, energy absorption refers to the amount of energy that is not reflected by the ABH, which is not only induced by resistive (dissipative) phenomena, but also, and especially, by reactive ones.}
For each element $i$, the form of $\mathbf{T}_i$ includes wave propagation through a straight duct, together with factors accounting for variations in the local acoustic impedance due to the presence of a perforated boundary and a side cavity. Under the 1D plane-wave approximation, and using the time convention $e^{j\omega t}$, the transfer matrix $\mathbf{T}_i$ is expressed as:

\begin{align}
\mathbf{T}_i = &
\begin{bmatrix}
\cos(k \Delta x/2) & -jZ_0 \sin(k\Delta x/2) \\
-\frac{j}{Z_0} \sin(k\Delta x/2) & \cos(k\Delta x/2)
\end{bmatrix} 
\begin{bmatrix}
1 & 0 \\
-\frac{S_{cav}/S_{duct}}{Z_{br_i}} & 1
\end{bmatrix} 
\begin{bmatrix}
\cos(k \Delta x/2) & -jZ_0 \sin(k \Delta x/2) \\
-\frac{j}{Z_0} \sin(k \Delta x/2) & \cos(k \Delta x/2) \label{eq:TMM_element}
\end{bmatrix}
\end{align}

Here, $\Delta x$ is the axial separation distance between the holes, $k$ is the wave number, $Z_0=\rho_0 c_0$ is the characteristic impedance of air, $Z_{br}$ is the impedance of the side branch, $S_{cav}$ is the surface area of the back cavity just after the perforated back plate, $S_{duct}$ is the cross section area of the duct. The side branch impedance is composed of two parts: the surface impedance of the perforated boundary, $Z_{pb}$, and the cavity impedance behind the \rx{perforator}\rn{perforated plate}, $Z_{c_i}$.
\noindent
Hence,
\[
Z_{br_i} \;=\; Z_{c_i} \;+\; Z_{\mathrm{PB}}.
\]
The cavity impedance backed by a rigid wall $Z_{c_i}$ is:
\begin{equation}
    Z_{c_i} = -jZ_0\cot(kH_i)\label{eq:Zci_complete},
\end{equation}
where $H_i$ is the height of the $i^\mathrm{th}$ cavity. When $k H_i \ll1$, the impedance above reduces to:
\begin{equation}
    Z_{c_i} \approx -j\frac{Z_0}{kH_i}=\frac{1}{j\omega \underbrace{\frac{H_i}{\rho_0c_0^2}}_{C_i}}\label{eq:lowfreq_Zci}
\end{equation}

Because the diameter of the \rx{perforator}\rn{perforated plate} holes of our damper is in the millimeter \rx{regime}\rn{range}, we adopt the Beranek--Ingard's model for the surface impedance of the \rx{perforator}\rn{perforated plate} \cite{BaranekIngard}. The model contains the \rx{resistive}\rn{resistance} $R$, additional viscothermal reactance $X_{VT}$, and inertance $M$ part of the impedance as:

\begin{equation}
Z_{\mathrm{PB{-}BI}} = \underbrace{\Bigl(\frac{4t_h}{d_\mathrm{h}}+4\Bigr) \, \frac{R_\mathrm{s}}{\delta}}_{R(\sqrt{\omega})} +  j\underbrace{\Bigl(\frac{4t_h}{d_\mathrm{h}}+4\Bigr) \, \frac{R_\mathrm{s}}{\delta}}_{X_{VT}(\sqrt{\omega})}+j\omega \, \underbrace{\frac{ \rho_0}{\delta} (\overbrace{2\,\epsilon_\mathrm{e{-}BI} + t_h}^{L_\text{eff}})}_{M} 
\label{eq:ZPB_BI}
\end{equation}
where $\delta$ is the perforation ratio, $t_h$ is the plate thickness, $d_\mathrm{h}$ the hole diameter, $L_\text{eff}$ is the effective depth of the hole, $R_\mathrm{s} = \frac{1}{2}\sqrt{2\eta\omega\rho_0}$ is the surface resistance capturing dissipative effects, with $\eta$ being the dynamic viscosity of air. The term $\epsilon_\mathrm{e{-}BI}$ is the additional length which is given by:

\begin{equation}
\epsilon_\mathrm{e{-}BI}
=
0.48 \, \sqrt{
\pi\Bigl(\frac{d_\mathrm{h}}{2}\Bigr)^2
\,}
\Bigl(
\,1 - 1.47 \,\sqrt{\delta} \;+\;0.47 \,\sqrt{\delta^3} 
\Bigr)
\;.
\label{eq:epsilon_BI}
\end{equation}
The real part of $Z_{\mathrm{PB{-}BI}}$ represents the acoustic resistance, governing energy dissipation due to viscous effects. Meanwhile, the imaginary part that is proportional to $\omega$ reflects the acoustic reactance, indicating inertia associated with fluid motion through the perforations. 

\rev{In contrast to the approach of \cite{Maury2025}, millimetre-scale perforations are adopted here instead of sub-millimetre apertures in order to maintain linear acoustic behavior under the high excitation levels relevant to thermoacoustic instabilities. Micro-\rx{perforator}\rn{perforated plate}s have higher acoustic resistance because \rx{it}\rn{they} operate\rx{s} in the regime where the acoustic boundary layer $\delta_{ac}=\sqrt{2\nu/\omega}$ is in the order of the hole radius $r_h$, thereby increasing the acoustic losses through viscous effect. However, it is well established that \rx{micro-}perforated panels \rn{can} exhibit amplitude-dependent (nonlinear) impedance at elevated sound pressure levels due to flow separation, jetting, and vortex shedding at the aperture edges \cite{Maa1998,temiz_2016,MELLING1973}. \rn{The nonlinearity of the acoustic response of the aperture depends also on the presence of bias and grazing flow, and new models were recently proposed in  \cite{Stoychev2024a,Stoychev2024b}.} Th\rx{is}\rn{e} nonlinear behavior is typically triggered when the acoustic particle displacement $\xi$ becomes comparable to the hole radius $r_h$ \cite{temiz_2016}. For a given acoustic velocity or pressure amplitude, this condition is therefore reached at much lower excitation levels in sub-millimetre perforations than in millimetre-scale apertures, causing micro-perforated panels to enter the nonlinear regime prematurely and leading to amplitude-dependent and less predictable damping characteristics. \rn{In the case of combustor application, acoustic nonlinearities of damper apertures can lead to complete failure of thermoacoustic instability control \cite{Miniero2023}}.}

\rev{In addition, the presence of a bias (purge) flow, which is often required in combustion systems, further amplifies shear-layer and jetting losses at the aperture edges, an effect that scales inversely with the hole diameter and porosity\rn{, e.g. }  \cite{SCARPATO2012,LAHIRI2017,Stoychev2024a,HUMBERT2025}. Consequently, micro-perforated panels are particularly sensitive to bias-flow-induced resistance increases and performance degradation, whereas millimetre-scale perforations exhibit significantly greater robustness \rn{for frequency ranges that are relevant for thermoacoustic instability control in combustors}. Moreover, in additively manufactured components, surface roughness and geometric tolerances represent a non-negligible fraction of the aperture size in the sub-millimetre regime, leading to large variability in the effective impedance. In contrast, millimetre-scale perforations are much less sensitive to such imperfections, providing a more repeatable and predictable acoustic response. For these reasons, millimetre-scale perforations are selected in the present study to ensure stable and robust damping performance under realistic thermoacoustic operating conditions. The choice of hole diameter in our study aligns with the \rn{ones} \rx{typical diameter} used \rx{in} \rn{for some high frequency Helmholtz dampers implemented in} practical combustor\rn{s, e.g. }\rx{liner and Hemholtz damper} \cite{schnell2009,Noiray20122753}}.


By combining eqs. \eqref{eq:Zci_complete}, \eqref{eq:lowfreq_Zci} \rn{and} \eqref{eq:ZPB_BI}, the side-branch impedance, $Z_{br_i}$ \rx{can be}\rn{is} written as:
\begin{align}
    Z_{br_i} &= R(\sqrt{\omega})+jX_{VT}(\sqrt{\omega})+j\omega M-jZ_0\cot(kH_i)\label{eq:Zbranch_complete}\\ 
    &\approx R(\sqrt{\omega})+jX_{VT}(\sqrt{\omega})+j\omega M+\frac{1}{j\omega C_i},
    \qquad kH_i \ll 1. \label{eq:Zbranch_lowfreq}
\end{align}
By neglecting the additional reactance $X_{VT}$, the resonance condition occurs at a frequency $\omega_{HR_i}$ where the inertance and compliance term sum up to zero:

\begin{equation}
    \omega_{HR_i} = \frac{1}{MC_i}= c_0 \sqrt{\frac{\delta}{L_\text{eff}H_i}} =c_0\sqrt{\frac{A_\text{holes}}{L_\text{eff}V_\mathrm{cav_i}}}\label{eq:HHresonance}
\end{equation}
where $A_\text{holes}$ and $V_\mathrm{cav_i}$ are the total area of the \rx{perforator}\rn{perforated plate} holes in each cell and the volume of the $\mathrm{i^{th}}$ cavity, respectively. Note that \rn{in} eq.\eqref{eq:HHresonance}\rn{, $\omega_{HR_i} $} is the Helmholtz resonance of the $\mathrm{i^{th}}$ cavity. At this condition, the absorption is high due to very high acoustic velocity fluctuations which will then be dissipated through the resistive part of the impedance.

To account for additional dissipative effects, a complex wavenumber and a complex acoustic impedance is introduced following the approach of \cite{BROOKE2020}. Viscous and thermal losses in the cavity and the duct are modeled using the Johnson–Champoux–Allard–Lafarge (JCAL) equivalent fluid formulation \cite{Johnson1987,Champoux1991,Lafarge1997}. Within this framwework , the effective density $\rho_e(\omega)$ and compressibility of air $C_{e}(\omega)$ are given by:

\begin{equation}
    \rho_e(\omega) = \rho_0\left(1-j\frac{12\eta}{\omega\rho_0w_{c,d}^2}\sqrt{1+j\frac{\omega\rho_0w_{c,d}^2}{36\eta}}\right), \label{eq:rho_e}
\end{equation}
\begin{equation}
    C_e(\omega) = \frac{1}{\rho_0c_0^2}\left(\gamma-\frac{\gamma-1}{1-j\frac{12\eta}{\text{Pr} \omega\rho_0 w_{c,d}^2}\sqrt{1+j\frac{\omega \rho_0 w_{c,d}^2\text{Pr}}{36\eta}}}\right), \label{eq:C_e}
\end{equation}
where $w_{c,d}$ denotes an effective thermo-viscous characteristic width of the cavity and the duct governing both the static airflow resistivity and the viscous and thermal characteristic frequencies entering the JCAL formulation, and Pr is the Prandtl number. The parameter $w_{c,d}$ should not be interpreted as the geometric spacing of the cavity or duct, but rather as a phenomenological length scale accounting for non-ideal viscous and thermal dissipation effects not captured by the idealized parallel-plate assumption. 

The complex wavenumber $k_c$ and the complex specific impedance $Z_c$ are expressed in terms of $\rho_e(\omega)$ and $C_e(\omega)$ as follows:

\begin{equation}
\begin{split}
    k_c(\omega) &= \omega\sqrt{\rho_e(\omega)C_e(\omega)},\\
    Z_c(\omega) &= \sqrt{\rho_e(\omega)/C_e(\omega)}.
\end{split}
\end{equation}

The $k$ and $Z_0$ in eq.~\eqref{eq:Zci_complete} are substituted by $k_c$ and $Z_c$, respectively. However, for the propagation of waves inside the duct of the ABH (the first and the last matrices in eq.~\eqref{eq:TMM_element}, we do not use the complex wave number formulation since the width of the duct and the frequency of interest are sufficiently large that thermo-viscous effects are negligible, and the approximations $k_c(\omega)\approx k_0$ and $Z_c(\omega) \approx Z_0$ are justified.

The scattering matrix is defined as: 

\begin{equation}
\begin{bmatrix} f_2 \\ g_1 \end{bmatrix}
=
\begin{bmatrix} S_{11} & S_{12} \\ S_{21} & S_{22} \end{bmatrix}
\begin{bmatrix} f_1 \\ g_2 \end{bmatrix}.
\end{equation}
The transfer matrix from eq.~\ref{eq:Global} can be converted into the scattering matrix by using the following transformations:

\begin{equation}
\begin{aligned}
\Gamma &= T_{11} - \frac{T_{12}}{Z_0} - T_{21} Z_0 + T_{22}, \\
S_{11} &= \frac{2\left(T_{11} T_{22} - \frac{T_{12}}{Z_0} \cdot T_{21} Z_0\right)}{\Gamma}, \\
S_{12} &= \frac{T_{11} - \frac{T_{12}}{Z_0} + T_{21} Z_0 - T_{22}}{\Gamma}, \\
S_{21} &= \frac{-T_{11} - \frac{T_{12}}{Z_0} + T_{21} Z_0 + T_{22}}{\Gamma}, \\
S_{22} &= \frac{2}{\Gamma}
\end{aligned}
\label{eq:scattering}
\end{equation}

Two dissipation coefficients are defined to characterize energy loss \rn{for incident waves originating from} upstream and downstream of the ABH:
\begin{equation}
    \begin{aligned}
        \alpha_u &= 1 - |S_{11}|^2 - |S_{21}|^2, \\
        \alpha_d &= 1 - |S_{12}|^2 - |S_{22}|^2,
    \end{aligned}
    \label{eq:dissipation}
\end{equation}
where $\alpha_u$ and $\alpha_d$ represent the upstream and downstream dissipation, respectively.

\subsection{Analytical treatment}

In order to analyze the perforated ABH damper analytically, we assume an infinite number of cavities with inifitesimally small width to ensure continous axial variation of the side branch impedance, following the approach in \cite{Mironov}. In this section, the effective density and the compressibility of air described in eqs.~\eqref{eq:rho_e} and \eqref{eq:C_e} are not employed. Assuming a one dimensional wave propagation, the conservation of mass can be written as:

\begin{equation}
    \begin{split}
    \frac{\partial\rho}{\partial t}S_\text{duct} dx + \rho_0 S_\text{duct}du+\rho_0u_{\perp}P_\text{cav}dx&=0\\
    \frac{\partial \rho}{\partial t}+\rho_0\frac{\partial u}{\partial x}+\rho_0u_{\perp}\underbrace{\frac{P_\text{cav}}{S_\text{duct}}}_{1/r_H}&=0,   
    \end{split}\label{eq:mass}
\end{equation}
where $P_\text{cav}$ is the perimeter of the side-branch cavity ($dS_\text{cav} = P_\text{cav} dx$), $r_H=P_\text{cav}/S_{duct}$ is the hydraulic radius, and $u_\perp = p/Z_\text{br}(x)$ is the normal velocity coming from the \rx{perforator}\rn{perforated plate}s. The \rx{euler}\rn{Euler} momentum equation remains intact:
\begin{equation}
    -\frac{\partial p}{\partial x}=\rho_0\frac{\partial u}{\partial x}.\label{eq:momentum}
\end{equation}
By combining eqs. \eqref{eq:mass} and \eqref{eq:momentum}, and using the pressure and density relation $c_0^2d\rho = dp$, the pressure wave equation can be written as follows:

\begin{equation}
    \frac{1}{c_0^2}\frac{\partial^2p}{\partial t^2}-\frac{\partial^2p}{\partial x^2}+\frac{\rho_0}{r_H Z_\text{br}(x)}\frac{\partial p}{\partial t} = 0.
\end{equation}
The corresponding \rx{pressure} Helmholtz equation \rn{for the acoustic pressure} is:

\begin{equation}
    \frac{\partial^2p}{\partial x^2} + \underbrace{\bigg[ \frac{\omega^2}{c_0^2}-\frac{i\omega\rho_0}{r_H Z_\text{br}(x)}\bigg]}_{k^2(x)} p= 0\label{eq:pressure-helmholtz}
\end{equation}
The terms inside the square bracket in eq.\eqref{eq:pressure-helmholtz} is defined as defined as the square of the axial wavenumber $k(x)$. By employing eq. \eqref{eq:Zbranch_lowfreq}, the following relation holds:
\begin{equation}
\begin{split}
    k^2(x,\omega) &= \frac{\omega^2}{c_0^2}\left(1-j\frac{\rho_0c_0}{\frac{\omega}{c}r_HZ_\text{br}(x)}\right)\\
    &= k_0^2\left(1-j\frac{Z_0}{k_0r_HZ_\text{br}(x)}\right)\\
  &= k_0^2\left(1-j\frac{Z_0}{k_0r_H} \left[R+jX_{VT} +j\omega M+\frac{1}{j\omega C(x)}\right]^{-1}\right)  
\end{split}
\end{equation}
In order to simplify the analysis, let us neglect $R$ and $X_{VT}$, and the following expression for the wave number $k(x,\omega)$ is obtained:
\begin{equation}
\begin{split}
    k(x,\omega) &= k_0\sqrt{1-j\frac{Z_0}{k_0r_H} \left[j\omega M+\frac{1}{j\omega C(x)}\right]^{-1}},\\
    &= k_0\sqrt{1-j\frac{Z_0}{k_0r_H} \left[\frac{j\omega C(x)}{1-\omega^2MC(x)}\right]},\\
    &= k_0\sqrt{1+\frac{H(x)}{r_H}\left[\frac{1}{1-\omega^2/\omega^2_{HR}(x)}\right]},
    \end{split}
\end{equation}
\rn{where $H(x)=\rho_0  c_0^2 \,C(x)$.} Consequently, the axial phase speed, $c_{ph}$ is:
\begin{equation}
    c_{ph}(x,\omega) = \frac{\omega}{c} = \frac{c_0}{\sqrt{1+\dfrac{H(x)}{r_H}\left[\dfrac{1}{1-\omega^2/\omega^2_{HR}(x)}\right]}}.\label{eq:phasespeed_nodamping}
\end{equation}

Therefore, at low frequencies, below the resonance of the deepest Helmholtz resonator, the axial phase speed can be written as:

\begin{equation}
    c_{ph}(x)\, \rn{\simeq}\rx{=} \,\frac{c_0}{\sqrt{1+\frac{H(x)}{r_H}}}, \quad \rn{\text{for}} \quad \omega/\omega_{HR}\ll 1.\label{eq:phasespeed_nodamping_lowfreq}
\end{equation}
Note that the group velocity, $c_g = (dk/d\omega)^{-1}$, coincides with the phase velocity $c_{ph}$ under the low-frequency approximation given in Eq.~\eqref{eq:phasespeed_nodamping_lowfreq}. This slow-sound effect is the same as the ABH presented in \cite{Bravo2023,GUASCH2017,SERRA2023,Martin2025}. In this regime, the compliance term dominates, making the denominator in Eq.~\eqref{eq:phasespeed_nodamping_lowfreq} greater than unity and thus reducing the effective speed of sound below $c_0$. Because $H(x)$ increases monotonically with $x$, the speed of sound correspondingly decreases monotonically along the longitudinal direction.
When, $\omega = \omega_{HR}(x)$, the denominator in eq.\eqref{eq:phasespeed_nodamping} blows up and $c_{ph}$ goes to zero. This corresponds to the zero transmission condition where the propagating wave is \rn{fully} reflected \rn{in the lossless scenario}. However, in reality, resistive effect always exist and so far, the resistive part $R(\sqrt{\omega})$ in eq.~\eqref{eq:ZPB_BI} has been neglected. The effect of \rx{the} $R(\sqrt{w})$ on the wave number $k(x,\omega)$ will be discussed later. 

When $\omega > \omega_{HR}(x)$, the denominator in eq.~\eqref{eq:phasespeed_nodamping} is less than unity and hence, the phase speed is larger than one. In this regime, the inertance part of the impedance dominates which in turns lead to supersonic phase speed. However, note that $c_g$ differs from $c_{ph}$. Substituting $\Omega = \Omega(x,\omega)^2 = \omega^2/\omega_{HR}(x)^2,~B = B(x) = H(x)/r_H$, the expression for $c_g$ reads:

\begin{equation}
    c_g(x,\omega) = \left(\frac{dk}{d\omega}\right)^{-1} 
    = c_0\frac{\sqrt{1+B\frac{1}{1-\Omega^2}}}{1+B\frac{1}{1-\Omega^2}+B\frac{\Omega^2}{(1-\Omega)^2}} = c_0\frac{\sqrt{1+\frac{B}{1-\Omega^2}}}{1+\frac{B}{(1-\Omega^2)^2}}. \label{eq:cg_nodamping}
\end{equation}

The denominator in Eq.~\eqref{eq:cg_nodamping} is always greater than unity and larger than the numerator, such that the group velocity satisfies $c_g(x,\omega)\le c_0$ whenever the wavenumber is real such that the information causality is always guaranteed. There exists a finite frequency interval in which the argument of the square root in Eq.~\eqref{eq:cg_nodamping} becomes negative, namely for $1<\Omega<\sqrt{1+B}$. In this interval, the axial wavenumber is purely imaginary and the acoustic field is evanescent, so that no propagating mode exists. This frequency range corresponds to a \rn{`}locally resonant acoustic band gap\rn{'} induced by the side-loaded Helmholtz resonator. At sufficiently high frequencies, $\Omega\to\infty$, the influence of the resonator vanishes and the group velocity asymptotically approaches $c_0$.

As mentioned previously, so far, we only consider the idealized treatment in which resistive effects are present. If the resistive part of the impedance, $R(\sqrt{\omega})$, is included, the expression for $k(x,\omega)$ reads:

\begin{equation}
    k(x,\omega) = k_0\sqrt{1+\frac{H(x)}{r_H}\left[\frac{1}{1+j\omega R(\sqrt{\omega})C(x)-\omega^2/\omega_{HR}(x)^2}\right]}.
\end{equation}
Hence, the wavenumber is complex and sound attenuation always persists for all $\omega \in \Re$. The maximum attenuation occurs at $\omega\approx\omega_{HR}(x)$, and since the Helmholtz frequency varies spatially, different frequency components interact resonantly at different axial locations. As a result, the locally resonant band gap of a uniform system is replaced by a spatially distributed high-attenuation region, leading to broadband acoustic attenuation.

Therefore, the device operates by slowing sound propagation at low frequencies and by introducing a spatially graded Helmholtz resonance that promotes distributed resonant dissipation, thereby enabling effective broadband acoustic absorption. Therefore, at low frequencies, the device behaves similarly as a side-mounted acoustic black hole as in \cite{Bravo2023}, while at higher frequencies it behaves similarly as rainbow trapping silencers as in \cite{Jimenez2017,Maury2025}. In some studies with similar design principles, such a device is called acoustic black hole with perforated coating or boundary \cite{Li2024,ZHU2025,li_slow_2025,Maury2025Perforated}

\section{Experimental setup}
\begin{figure}[b!]
    \centering
    \begin{subfigure}{0.6\textwidth}
    \def\svgwidth{1\textwidth}
\begingroup%
  \makeatletter%
  \providecommand\color[2][]{%
    \errmessage{(Inkscape) Color is used for the text in Inkscape, but the package 'color.sty' is not loaded}%
    \renewcommand\color[2][]{}%
  }%
  \providecommand\transparent[1]{%
    \errmessage{(Inkscape) Transparency is used (non-zero) for the text in Inkscape, but the package 'transparent.sty' is not loaded}%
    \renewcommand\transparent[1]{}%
  }%
  \providecommand\rotatebox[2]{#2}%
  \newcommand*\fsize{\dimexpr\f@size pt\relax}%
  \newcommand*\lineheight[1]{\fontsize{\fsize}{#1\fsize}\selectfont}%
  \ifx\svgwidth\undefined%
    \setlength{\unitlength}{514.54472099bp}%
    \ifx\svgscale\undefined%
      \relax%
    \else%
      \setlength{\unitlength}{\unitlength * \real{\svgscale}}%
    \fi%
  \else%
    \setlength{\unitlength}{\svgwidth}%
  \fi%
  \global\let\svgwidth\undefined%
  \global\let\svgscale\undefined%
  \makeatother%
  \footnotesize
  \begin{picture}(1,0.43597561)%
    \lineheight{1}%
    \setlength\tabcolsep{0pt}%
    \put(0,0){\includegraphics[width=\unitlength,page=1]{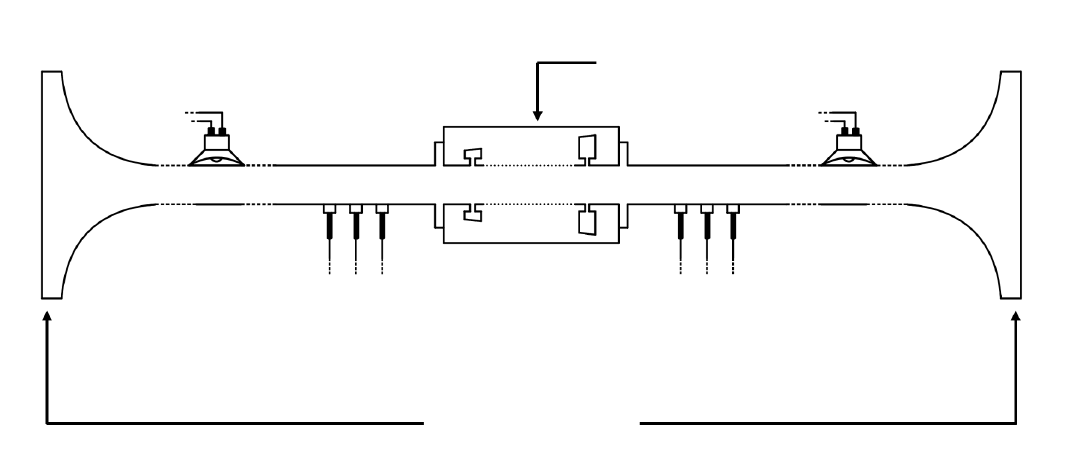}}%
    \put(0.13607199,0.3831236){\color[rgb]{1,1,1}\makebox(0,0)[lt]{\lineheight{1.25}\smash{\begin{tabular}[t]{l}L\end{tabular}}}}%
    \put(0.13,0.34561273){\color[rgb]{0,0,0}\makebox(0,0)[lt]{\lineheight{1.25}\smash{\begin{tabular}[t]{l}Loudspeaker\end{tabular}}}}%
    \put(0.7,0.34584607){\color[rgb]{0,0,0}\makebox(0,0)[lt]{\lineheight{1.25}\smash{\begin{tabular}[t]{l}Loudspeaker\end{tabular}}}}%
    \put(0.55,0.14){\color[rgb]{0,0,0}\makebox(0,0)[lt]{\lineheight{1.25}\smash{\begin{tabular}[t]{l}Microphones\end{tabular}}}}%
    \put(0.24,0.14){\color[rgb]{0,0,0}\makebox(0,0)[lt]{\lineheight{1.25}\smash{\begin{tabular}[t]{l}Microphones\end{tabular}}}}%
    \put(0.41,0.04564093){\color[rgb]{0,0,0}\makebox(0,0)[lt]{\lineheight{1.25}\smash{\begin{tabular}[t]{l}Anechoic \\termination\end{tabular}}}}%
    \put(0.55979566,0.37377614){\color[rgb]{0,0,0}\makebox(0,0)[lt]{\lineheight{1.25}\smash{\begin{tabular}[t]{l}ABH\end{tabular}}}}%
  \end{picture}%
\endgroup%

    \end{subfigure}
    \hfill
    \hspace{-2.5em}
    \begin{subfigure}{0.34\textwidth}
    \def\svgwidth{1\textwidth}
\begingroup%
  \makeatletter%
  \providecommand\color[2][]{%
    \errmessage{(Inkscape) Color is used for the text in Inkscape, but the package 'color.sty' is not loaded}%
    \renewcommand\color[2][]{}%
  }%
  \providecommand\transparent[1]{%
    \errmessage{(Inkscape) Transparency is used (non-zero) for the text in Inkscape, but the package 'transparent.sty' is not loaded}%
    \renewcommand\transparent[1]{}%
  }%
  \providecommand\rotatebox[2]{#2}%
  \newcommand*\fsize{\dimexpr\f@size pt\relax}%
  \newcommand*\lineheight[1]{\fontsize{\fsize}{#1\fsize}\selectfont}%
  \ifx\svgwidth\undefined%
    \setlength{\unitlength}{317.31333575bp}%
    \ifx\svgscale\undefined%
      \relax%
    \else%
      \setlength{\unitlength}{\unitlength * \real{\svgscale}}%
    \fi%
  \else%
    \setlength{\unitlength}{\svgwidth}%
  \fi%
  \global\let\svgwidth\undefined%
  \global\let\svgscale\undefined%
  \makeatother%
  \begin{picture}(1,0.78138461)%
  \footnotesize
    \lineheight{1}%
    \setlength\tabcolsep{0pt}%
    \put(0,0){\includegraphics[width=\unitlength,page=1]{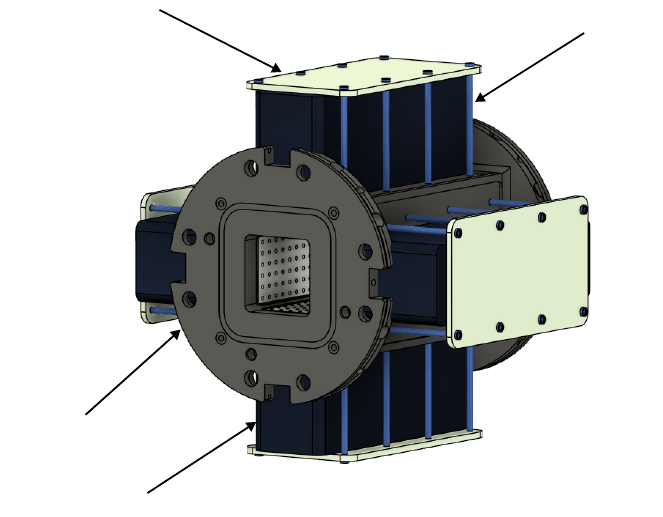}}%
    \put(-0.06,0.11577631){\color[rgb]{0,0,0}\makebox(0,0)[lt]{\lineheight{1.25}\smash{\begin{tabular}[t]{l}Cage module\end{tabular}}}}%
    \put(0.17776921,-0.01){\color[rgb]{0,0,0}\makebox(0,0)[lt]{\lineheight{1.25}\smash{\begin{tabular}[t]{l}ABH\end{tabular}}}}%
    \put(0.15347251,0.77){\color[rgb]{0,0,0}\makebox(0,0)[lt]{\lineheight{1.25}\smash{\begin{tabular}[t]{l}Plate\end{tabular}}}}%
    \put(0.87689102,0.75005617){\color[rgb]{0,0,0}\makebox(0,0)[lt]{\lineheight{1.25}\smash{\begin{tabular}[t]{l}Fastening\\   screw\end{tabular}}}}%
  \end{picture}%
\endgroup%

    \end{subfigure}
    \caption{a) Non-reactive test rig. b) Perforated ABH damper mounted on the cage module of the combustor setup. The damper is aligned such that the cavity height increases from left to right.}
    \label{fig:setup_nonreactive_ABH}
\end{figure}
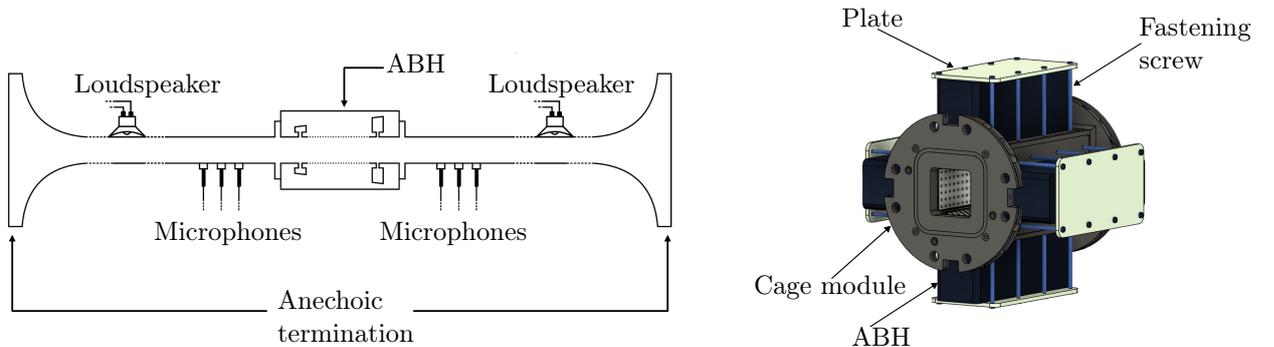

The non-reactive experimental setup for the measurements of the scattering matrix of the ABH designs is shown in Fig.~\ref{fig:setup_nonreactive_ABH}a). The setup consists of anechoic terminations, rectangular duct with 62 $\times$ 62~$\mathrm{mm^2}$ cross section, loudspeakers, and three microphones on each side of the test rig. Multi microphone method (MMM) is used to reconstruct the Riemann invariant component upstream and downstream of the ABH and subsequently the scattering matrix is recovered. 

Four units of each ABH design described in the next section are 3D printed and mounted on the four sides of the combustor cage module, as shown in Fig.~\ref{fig:setup_nonreactive_ABH}b). The combustor cage module also has a cross section area of 62$\times$62~$\mathrm{mm^2}$. To secure the ABH, four flat plates with eight holes each are used in combination with fastening screws. The cage module is compatible with both non-reactive and reactive test rigs in our laboratory.

\begin{figure*}
    \centering
    \makebox[\textwidth][c]{%
        \def\svgwidth{1.2\textwidth}
        \input{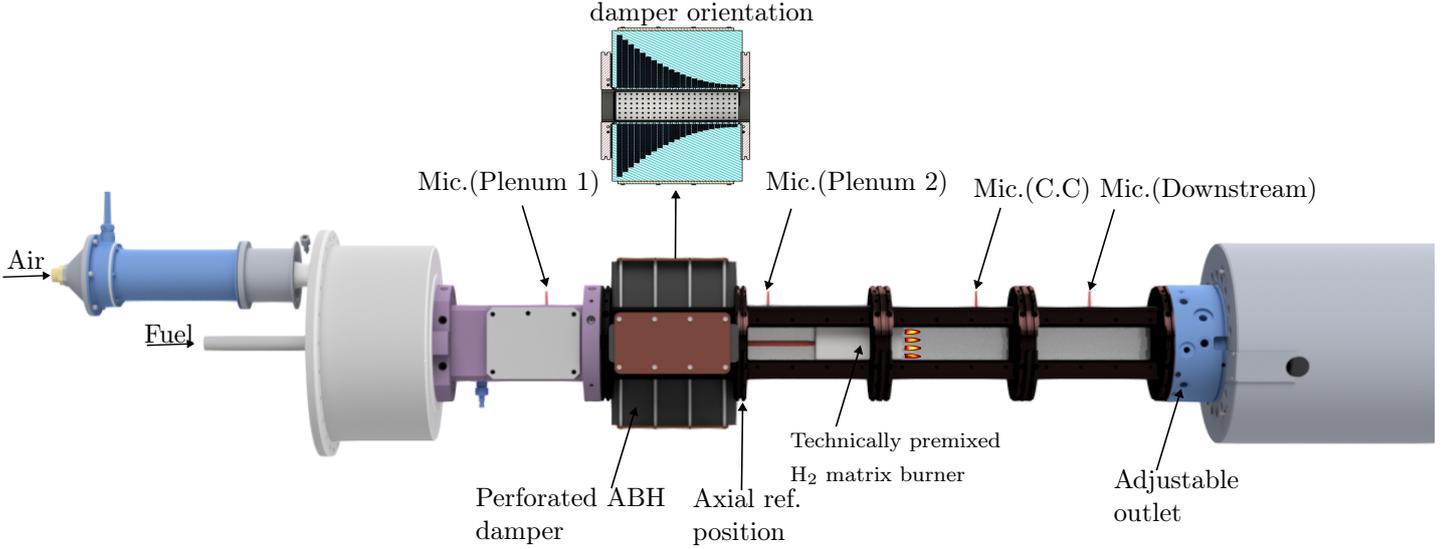}
    }
    \vspace{0.3cm}
    \caption{Lab-scale atmospheric low Mach number combustor equipped with microphones and perforated ABH damper. Mic: microphones, ABH: Acoustic black hole, C.C: combustion chamber. For the baseline configuration, the ABH blocks are replaced with glass windows.}
    \label{fig:setup_reactive}
\end{figure*}

The reactive test rig is shown in Fig.~\ref{fig:setup_reactive}. The setup is composed of multiple cage modules with a cross section of 62$\times$62 $\mathrm{mm^2}$ and a length of 250~mm. Two microphones, one upstream and one dowsntream of the ABH, are mounted on the plenum side, labeled as Plenum~1 and Plenum~2, respectively. Two microphones are mounted downstream of the burner, one microphone is measuring the pressure pulsation inside the combustion chamber, and another one downstream of the combustion chamber. The ABH is mounted on each side of one of the upstream cage modules. The outlet boundary is equipped with a motor-driven adjustable water-cooled piston as described in \cite{Dharmaputra2023,Blonde2025}. Depending on operating conditions and instability frequencies, the adjustable outlet area can change the system's stability effectively. Changing the outlet area effectively modifies the outlet reflection coefficients of the setup. The burner is a technically premixed 4$\times$ 4 hydrogen matrix burner, operated at a thermal power, $P_{th}$, of approximately 30~kW. The hydrogen fuel is injected through a fuel lance and ejected out of small holes inside the matrix channel, whereas the air is injected directly into the plenum section. Details about the matrix burner can be found in \cite{Moon2024}. The air mass flow rate is kept at about 17.5~g/s, corresponding to a mean bulk velocity of 3.8~m/s in the plenum duct section and 40~m/s in each matrix channel. The resulting relative pressure drop across the burner, $\Delta p/p$, is 1.6\%. The Mach number is 0.012 in the plenum section and about 0.03 downstream of the flame.

Three equivalence ratios are investigated: 0.475, 0.5, and 0.525. For all conditions, the outlet area is progressively swept from 13~$\mathrm{cm^2}$ to 18~$\mathrm{cm^2}$, each with a measurement duration of 30~s. The flame $\mathrm{OH^*}$ chemiluminescence is recorded with a highspeed intensified camera (Photron SA-X2 +LaVision HS IRO X) equipped with a UV lens (CERCO 100~mm f/2.8) and a narrowband bandpass filter (Edmund Optics 310~nm $\pm$ 10~nm) at a frame rate of 5~kHz. The effectiveness of the perforated ABH damper in suppressing thermoacoustic instabilities is assessed by comparing pressure pulsations with and without the perforated ABH damper. For the baseline configuration, the dampers are replaced with glass windows, resulting in a perfect 1D wave guide channel.

\begin{figure}
\centering
    \def\svgwidth{1\textwidth}
    \input{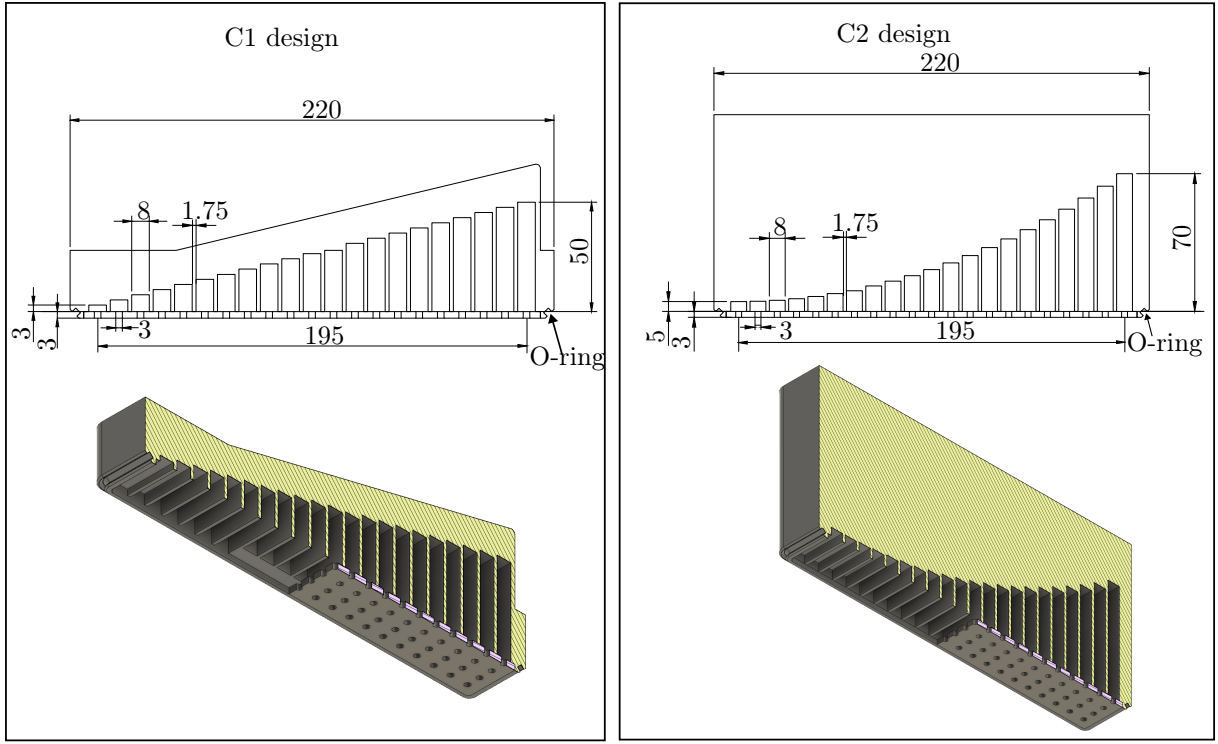}
    \caption{C1 (left) and C2 (right) damper design. The \rx{perforator}\rn{perforated plate} is printed separately and glued on to the bottom surface of the ABH block. Dimensions in mm.}
    \label{fig:c1_c2_design}
\end{figure}

\section{Results}

\subsection{Non-reactive test results}
Two perforated ABH damper configurations, denoted C1 and C2 (Fig.~\ref{fig:c1_c2_design}), are considered for model validation. The C1 design employs a linearly varying cavity height ($m=1$ in eq.~\ref{eq:profile_func}), with $H_{\min}=3$~mm and $H_{\max}=50$~mm. In contrast, the C2 design features a quadratic variation of the cavity height ($m=2$ in eq.~\ref{eq:profile_func}), with $H_{\min}=5$~mm and $H_{\max}=70$~mm. The cavity width is fixed at $w_c=8$~mm. In both designs, a perforated plate is located at the bottom of each cavity, with a hole diameter of $d_h=3$~mm, a pitch of 7~mm and thickness of $t_h = 3$~mm. Each cell contains seven holes, corresponding to a perforation ratio of $\delta=10\%$, and the axial spacing between successive holes is $\Delta x=9.75$~mm. The total axial length of each damper is fixed at 220~mm to ensure compatibility with the cage module shown in Fig.~\ref{fig:setup_nonreactive_ABH}b). 

Both designs are fabricated using a fused deposition modeling (FDM) 3D printer (Bambu Lab X1C) with ABS (Acrylonitrile Butadiene Styrene) material. For each configuration, four identical units are printed and installed on the four side walls of the combustor cage module, as shown in Fig.~\ref{fig:setup_nonreactive_ABH}b). The perforated plates are printed separately and subsequently bonded to the main damper body.

The scattering matrix is obtained by forcing the system from 100~Hz to 2500~Hz with a step of 25~Hz. The multi-microphone method (MMM) \cite{Paschereit2002,Schuermans2004} is employed to quantify the forward and backward propagating wave upstream and downstream of the damper, and subsequently the scattering matrix elements are computed. In the non-reactive test, the cavity height progressively increases from left to right.

\subsubsection{Scattering matrix measurements}\label{sec:SM_meas}
\begin{figure*}[t!]
    \centering
    \makebox[\textwidth][c]{%
        \def\svgwidth{1.2\textwidth}
        \input{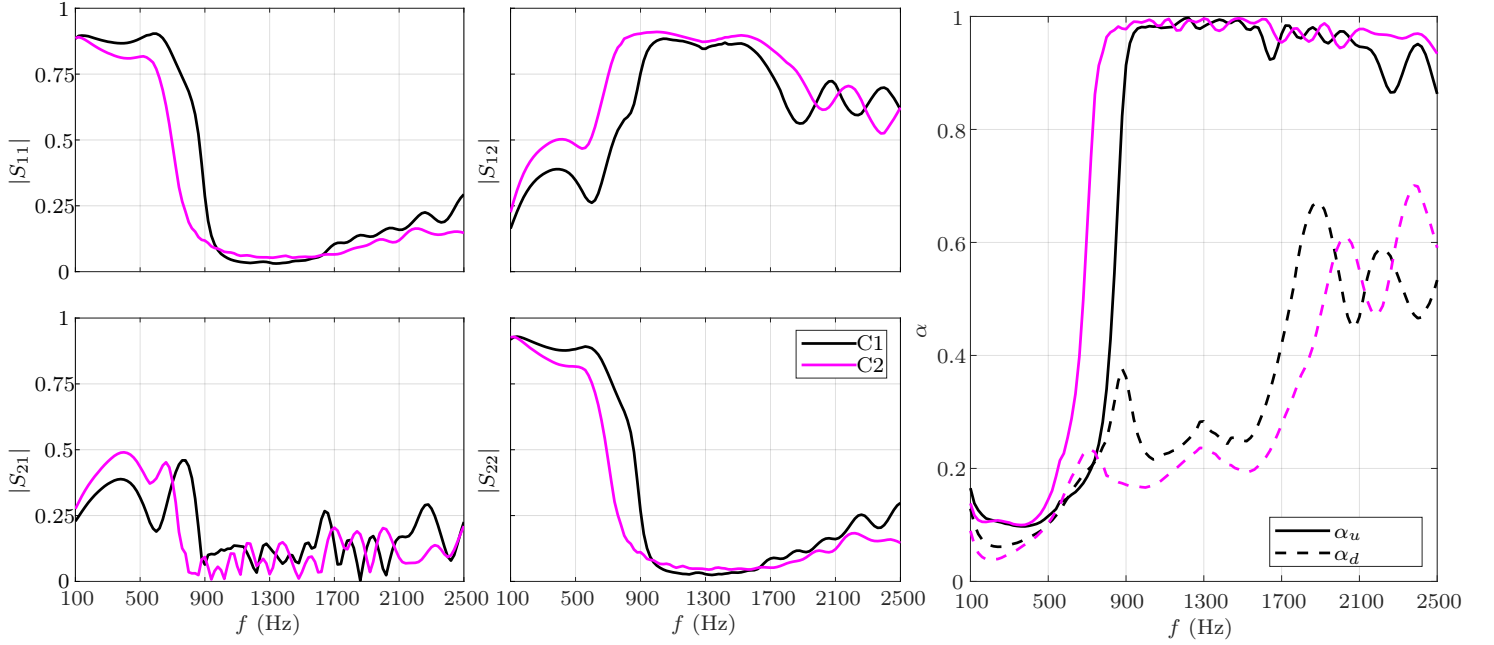}
    }
    \caption{Measured a) scattering matrix, b) dissipation coefficients of C1 and C2 designs. $\alpha_u:$ upstream dissipation, $\alpha_d:$ downstream dissipation. }
    \label{fig:c1_c2_exp}
\end{figure*}

Figure~\ref{fig:c1_c2_exp}a) presents the absolute values of the scattering matrix components of the  C1 and C2 design, obtained from measurements conducted in the non-reactive test rig. In this context, \( S_{11} \) and \( S_{22} \) represent the transmission coefficients for waves propagating from the upstream and downstream sides, respectively, while \( S_{21} \) and \( S_{12} \) correspond to the upstream and downstream reflection coefficients. The measurements indicate that, for both designs, the transmission coefficients \( S_{11} \) and \( S_{22} \) are almost identical across the investigated frequency range, suggesting a high degree of symmetry in the forward and backward transmission characteristics. Nevertheless, the frequency at which these coefficients begin to decay differs between the two designs: for C1, a noticeable drop in transmission occurs at approximately \(900~\text{Hz}\), whereas for C2, this drop occurs earlier, at around \(700~\text{Hz}\). 

In contrast, the reflection coefficients exhibit asymmetric behavior. Specifically, \( S_{21} \), corresponding to reflection from downstream to upstream, consistently shows a significantly lower gain compared to \( S_{12} \), which represents reflection from upstream to downstream. This asymmetry becomes particularly evident at higher frequencies, with the reduction in \( S_{21} \) beginning near \(900~\text{Hz}\) for C1 and \(700~\text{Hz}\) for C2\rx{---}\rn{,} closely matching the transmission drop-off points. These results suggest that while both configurations maintain comparable transmission performance in both directions ($S_{11}\approx S_{22}$), their reflection characteristics differ substantially, which can have important implications in practical applications as will be discussed later.



    

The dissipation coefficients for the two ABH designs, calculated using Eq.~\ref{eq:dissipation}, are presented in Fig.~\ref{fig:c1_c2_exp}b). The dissipation behavior is not symmetric for either configuration: in both cases, the \rx{upstream} dissipation coefficients \rn{for waves originating from the upstream side} are consistently higher than their downstream counterparts, $\alpha_u>\alpha_d$, which is caused by the lower value of $|S_{21}|$ compared to $|S_{12}|$. Furthermore, the frequency at which each design becomes effective differs notably. The cut-on frequency for the dissipation is defined when $\alpha_u(f) = 0.8$. For C1, the cut on frequency is \(900~\text{Hz}\), whereas for C2, this onset occurs earlier, near \(700~\text{Hz}\). This difference in the effective cut-on frequency can be attributed to the difference in $H_{max}$ between the two designs. The higher $H_{max}$ in C2 enables earlier onset of dissipative effects. These observations, when considered alongside the results of the scattering matrix, highlight the influence of geometric parameters on both reflection and dissipation performance.

\begin{figure*}[b!]
    \centering
    \makebox[\textwidth][c]{%
        \def\svgwidth{1.2\textwidth}
        \input{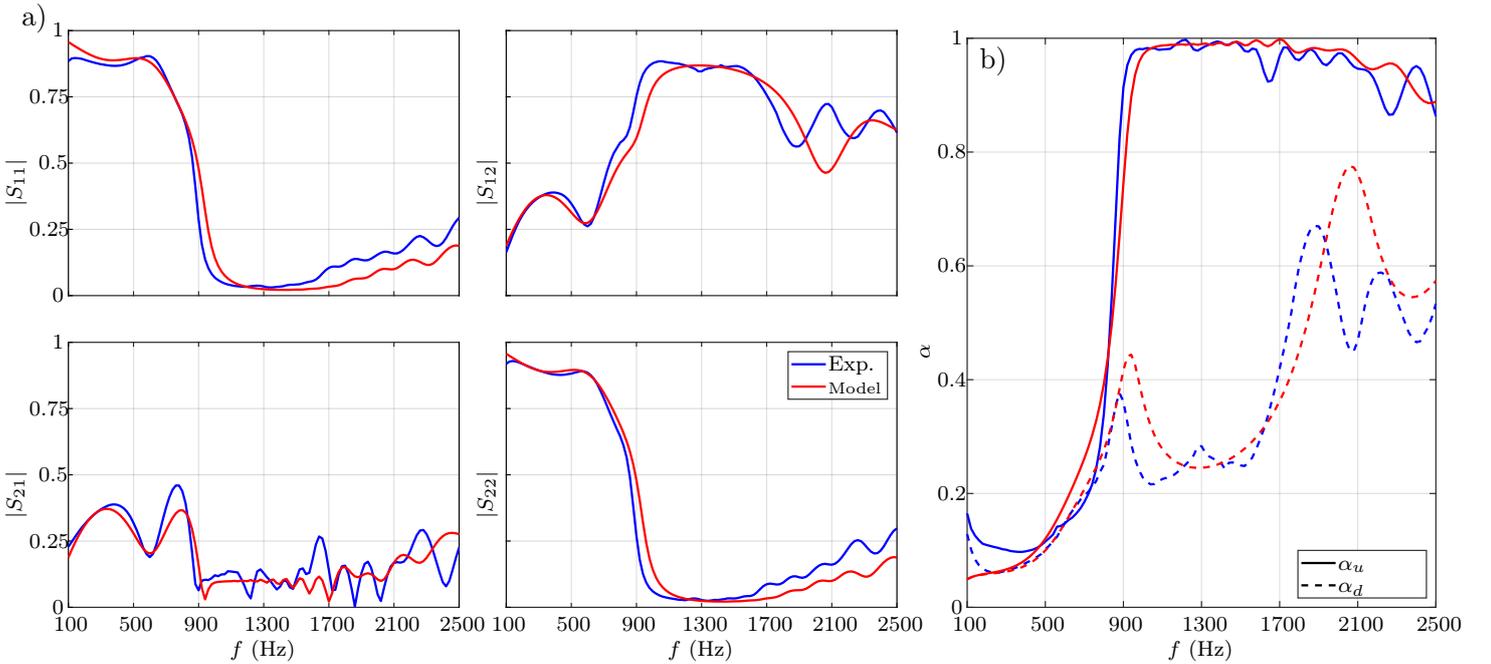}
    }
    \caption{Measured and modeled a) scattering matrix, b) dissipation coefficients of C1 design. $\alpha_u:$ upstream dissipation, $\alpha_d:$ downstream dissipation. }
    \label{fig:c1_exp_sim}
\end{figure*}

\begin{figure*}[t!]
    \centering
    \makebox[\textwidth][c]{%
        \def\svgwidth{1.2\textwidth}
        \input{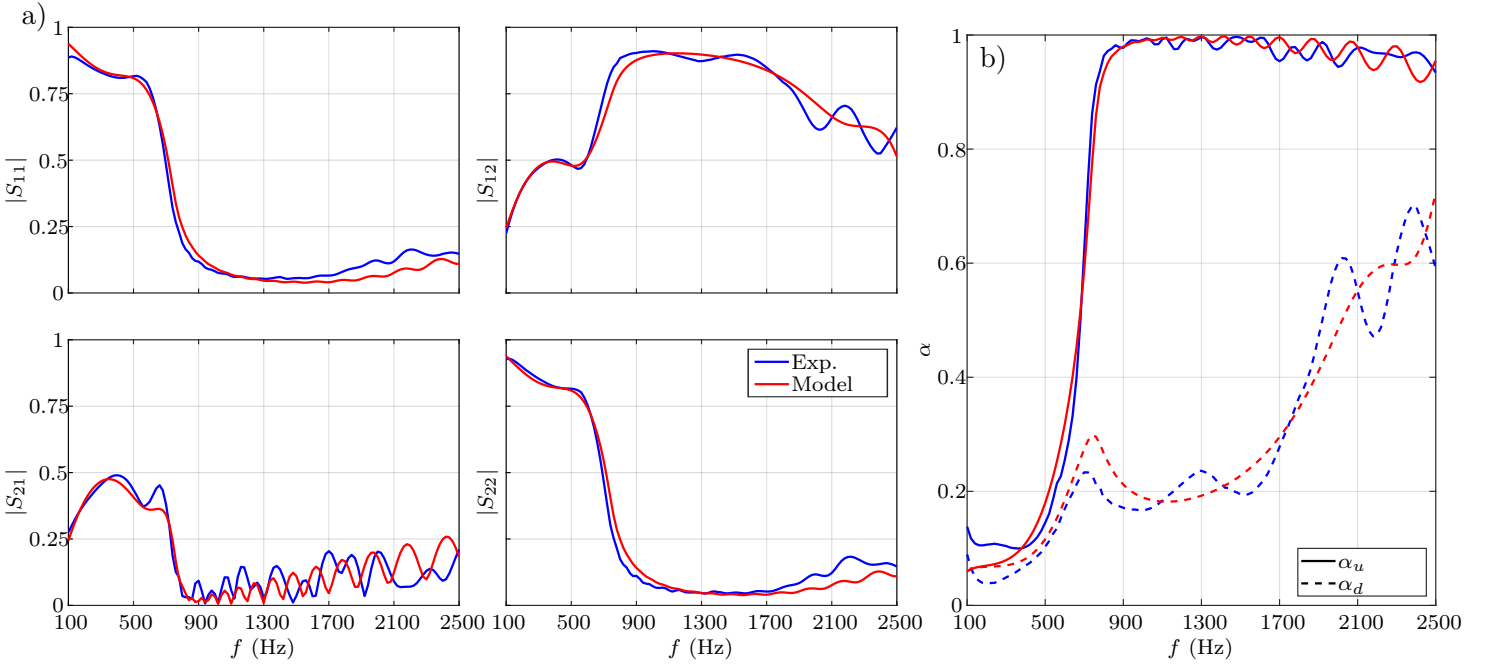}
    }
    \caption{Measured and modeled a) scattering matrix, b) dissipation coefficients of C2 designs. $\alpha_u:$ upstream dissipation, $\alpha_d:$ downstream dissipation.}
    \label{fig:c2_exp_sim}
\end{figure*}

The TMM outlined in the Sec.~\ref{sec:TMM} is evaluated with the geometrical parameters of C1 and C2. By using Eq.~\ref{eq:scattering}, the modeled transfer matrix is converted to the scattering matrix. The comparisons between TMM model and experiments of C1 and C2 designs are shown in Fig.~\ref{fig:c1_exp_sim} and Fig.~\ref{fig:c2_exp_sim}, respectively. The TMM model successfully captures the frequency response of each scattering matrix element. Some discrepancy occurs probably due to the precision of the 3D printer. However, the decay of transmission coefficients is successfully captured as well as the decay in the gain of $|S_{21}|$. This highlights the remarkable capability of the simple model in predicting the effective cut-on frequency of the ABH design.

Figures \ref{fig:c1_exp_sim}b) and \ref{fig:c2_exp_sim}b) compare the modeled and measured dissipation coefficients of the C1 and C2 design, respectively. Owing to the strong agreement between the modeled and measured scattering matrices, the dissipation coefficients also show good correspondence. Since Eq.~\ref{eq:dissipation} involves a squaring operator, even small discrepancies in the scattering matrices are amplified in the dissipation coefficients. Despite this sensitivity, the model accurately predicts the cut-on frequency of C1 and C2 This demonstrates that the simple model is suitable for further ABH design optimization, depending on the targeted effective cut-on frequency and the desired frequency bandwidth.

Using this information, a strategy can be devised to implement the perforated ABH damper in a combustion system setup. Because of the high preferential dissipation effect of the damper $\alpha_u>\alpha_d$, it can be arranged to maximally dissipate the acoustic perturbations from the flame. Therefore, if the damper is placed in the plenum, it must be oriented so that the depth of the back cavity decreases along the longitudinal direction and it must be flipped if it is mounted on the sections downstream of the burner. The direct reflection coefficient of the combustor setup (either upstream or downstream) will be governed by the $S_{21}$ in Fig.~\ref{fig:c1_c2_exp}a). Meanwhile, the other components will interact with the acoustical element adjacent to it.

Let us now consider the placement of the perforated ABH damper as in Fig.~\ref{fig:setup_reactive}, and let $R_u$ denote the reflection coefficient upstream of the ABH. The effective reflection coefficient $R_\mathrm{eff}$ at the axial reference location can be derived and result in the following:
\begin{equation}
    R_{\mathrm{eff}} = \frac{R_u S_{11}S_{22}}{1-R_uS_{12}} + S_{21}.
    \label{eq:effective_ref}
\end{equation}
The first term on the right-hand side corresponds to the interaction of the damper and the acoustical element adjacent to it. The second term corresponds to the ``Direct'' reflection of the ABH. Note that, assuming that the transmission coefficients are symmetric, the term $S_{11}S_{22}$ can be replaced by $S_{11}^2$. When $S_{11}\rightarrow0$, $R_{\mathrm{eff}} \rightarrow S_{21}$, therefore, minimizing both the $S_{11}$ and $S_{21}$ is important if one wants to minimize the effective reflection coefficient or in other words, maximizing the dissipation coefficients, $\alpha_u$. The placement of the damper can then be seen as modifying the boundary conditions of the system which can be an effective way to mitigate the thermoacoustic instabilities \cite{Blonde2025}. Modifications of the inlet or outlet reflection coefficients can lead to a decrease in the thermoacoustic gain potential, which can stabilize the system \cite{EMMERT2015}. \rev{It is not always guaranteed that reducing reflections through the boundaries lead to stabilization due to the intrinsic thermoacoustic instabilities (ITA), which is mainly dictated by the flame transfer function itself (FTF), rather than the acoustic eigenmodes of the system \cite{HOEIJMAKERS2014,EMMERT2015,EMMERT2017,SILVA2023}. Nevertheless, for the combustor configurations investigated in the present study, as well as in our previous experimental investigations, a consistent empirical correlation has been observed between a reduction of the inlet and outlet reflection coefficients and the suppression of thermoacoustic oscillations \cite{Blonde2025,DHARMAPUTRA2024_PROCI}. This is inline with the observations reported in the studies dealing with the pulsation reductions in industrial gas turbine combustors upon introducing acoustic dampers \cite{Bellucci2004_HHD,Bellucci2005,schnell2009,Noiray20122753,Bothien2013}. This suggests that, for the operating conditions considered here, the thermoacoustic behavior remains strongly influenced by boundary-mediated thermoacoustic feedback, such that modifying the effective boundary conditions through the installation of the perforated ABH damper constitutes an effective stabilization strategy.}



\subsubsection{Mean flow effects on the damper performance}
\begin{figure*}[t!]
    \centering
    \makebox[\textwidth][c]{%
        \def\svgwidth{1.2\textwidth}
        \input{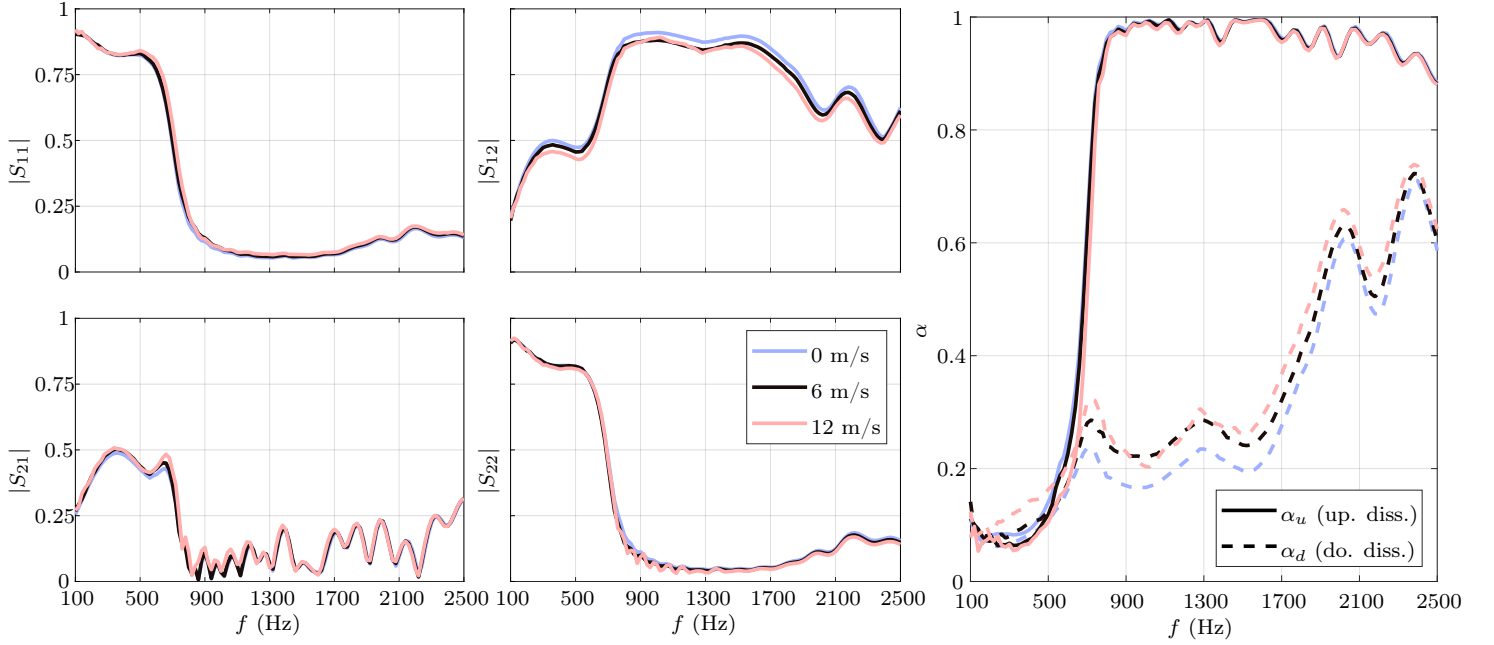}
    }
    \caption{Mean flow effects on the scattering and dissipation coefficients of C2 design. }
    \label{fig:c2_flow_effect}
\end{figure*}

The influence of duct mean flow on the performance of the perforated ABH damper is quantified, which is important since a non-zero mean flow is inevitably present in the combustor configuration. Figure~\ref{fig:c2_flow_effect} presents the scattering matrix and dissipation coefficients of the C2 design for three mean flow velocities: 0~m/s, 6~m/s, and 12~m/s. The maximum mean flow velocity achievable in the combustor setup in Fig.~\ref{fig:setup_reactive} is 7.5~m/s; therefore, the highest value considered here (12~m/s) exceeds the operational range of the combustor. Nevertheless, both the scattering matrix and the dissipation coefficients exhibit negligible sensitivity to the mean flow velocity. Consequently, the TMM model, which neglects mean flow effects, is sufficient to accurately predict the acoustic response of the perforated ABH damper. It is worth mentioning that in the intermediate mean flow velocities, we do not observe any flow-induced whistling. Furthermore, no measurable pressure drop penalty is observed when the damper is installed compared to the baseline configuration where it is replaced with glass windows.

\subsubsection{Profile effects of the ABH design}

The cavity height profile defined in Eq.~\ref{eq:pressure-helmholtz} corresponds to a \textit{convex} function, such that the geometric variation is primarily concentrated toward the outlet of the damper. This differs from the conventional ABH profile commonly adopted in the literature, which is \textit{concave} and therefore concentrates the grading near the inlet of the damper \cite{Bravo2023,Bravo2024,GUASCH2017,Maury2025}. Since the modeling framework has been experimentally validated in Sec.~\ref{sec:SM_meas}, the influence of the cavity height profile can be examined within the context of the present configuration.

For comparison, the conventional ABH profile, expressed in the present coordinate system, can be written as
\begin{equation}
    \hat{H}(x) = H_{min}+\left(H_{max}-H_{min}\right)\left[1-\left(1-\frac{x}{L_d}\right)^m\right],
    \label{eq:profile_func_ori}
\end{equation}
where $m$ denotes the grading exponent. Using the geometric parameters of configuration C2, the upstream dissipation coefficient $\alpha_u$ is computed for both the present profile (Eq.~\eqref{eq:profile_func}) and the conventional profile (Eq.~\eqref{eq:profile_func_ori}) over a range of exponent values.

\begin{figure}
    \centering
    \def\svgwidth{1\textwidth}
    \input{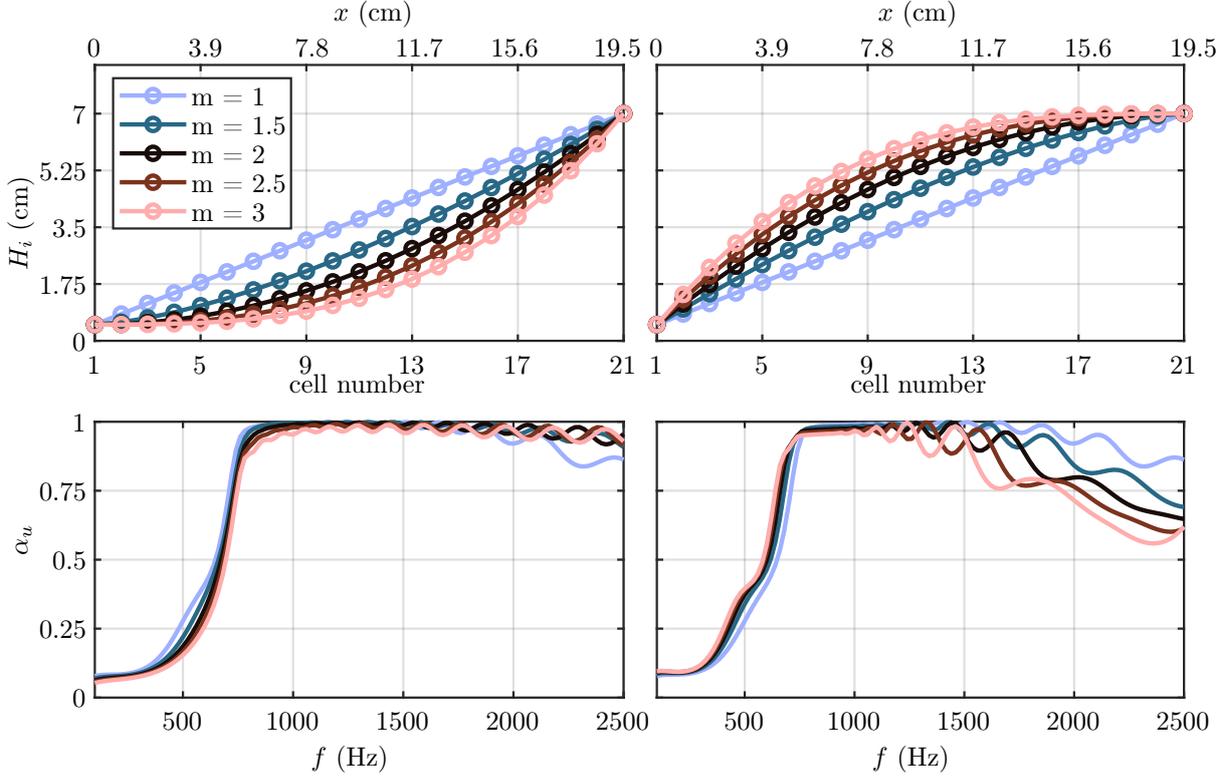}
    \caption{Effect of cavity height profile on the upstream dissipation coefficient. The cavity height profile and the upstream dissipation coefficient with profile function from eq.~\eqref{eq:profile_func} (left) and eq.~\eqref{eq:profile_func_ori} (right). The cavity height profile on the right corresponds to the conventional ABH profile.}
    \label{fig:profile_comp}
\end{figure}

Figure~\ref{fig:profile_comp} compares the resulting dissipation coefficients. Both profiles coincide for $m=1$, while differences emerge for $m>1$, particularly at frequencies above 1000~Hz. For the parameters considered here, the conventional profile exhibits a slightly lower cut-on frequency, which may be attributed to the larger proportion of cavities with greater height. However, its dissipation performance decreases at higher frequencies. In contrast, the present convex profile maintains higher dissipation levels over a broader frequency range. The convex grading may act as a progressive impedance transition that remains better matched at higher frequencies for the present geometry, hence, limiting the residual reflection and preserving high absorption. Notably, for the present geometry and perforation parameters, satisfactory performance is already obtained for $m=2$, which motivates the choice of this exponent in the remainder of this study.

\subsubsection{Refined ABH design}
\begin{figure*}
    \centering
    \def\svgwidth{1.2\textwidth}
    \input{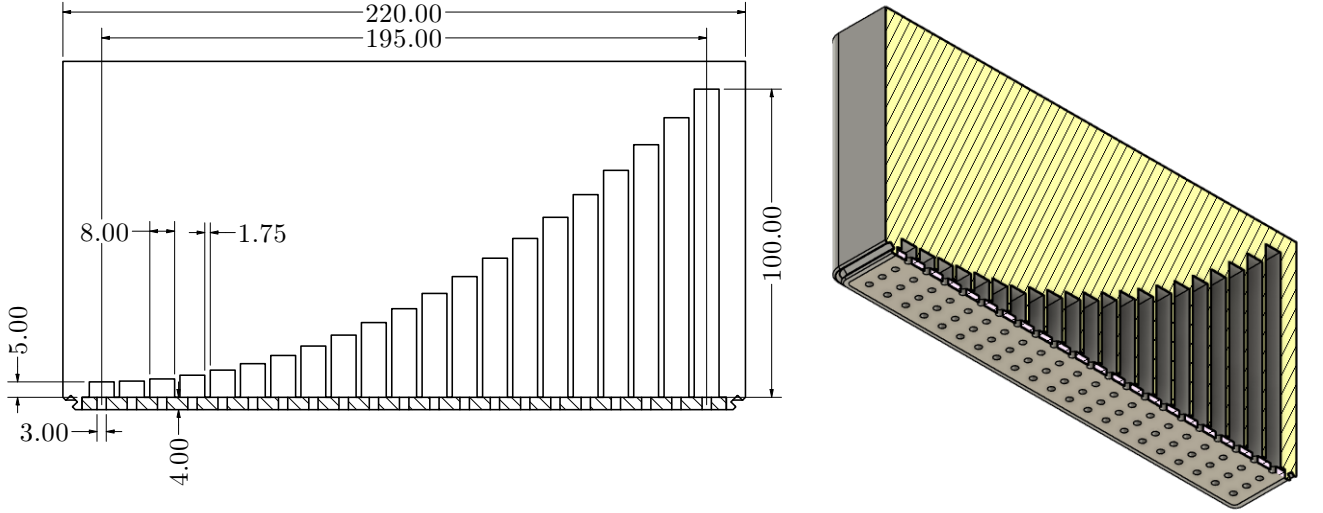}
    \caption{Optimized perforated ABH damper design (C3 design). Dimensions in mm.}
    \label{fig:C3_design}
\end{figure*}

Using the validated TMM model, a new design is devised by optimizing the $\alpha_u$ within the band of 500~Hz to 2000~Hz. This bandwidth is chosen because the typical instability frequencies of our combustor configuration operated with hydrogen is around 800-1200~Hz. The $H_{max}$ is set to 100~mm which is the maximum allowed height to be compatible with our lab-scale atmospheric and high pressure test-rig \rn{\cite{FAUREBEAULIEU2024,DHARMAPUTRA2024_PROCI}}. The holes diameter, $d_h$, and the \rx{perforator}\rn{perforated plate} thickness, $t_h$, are optimized as follows:

\begin{equation}
    \{d_h^*, t^*\} = \max_{d_h,t_h} \int_{500}^{2000} \alpha_u(f) ~df,
\end{equation}
The cavity height profile is follows Eq.~\ref{eq:profile_func} with $m=2$. The simple optimization leads to $d_h = 3~\mathrm{mm}, t_h=4~\mathrm{mm}$. The optimized design is labeled C3 design and shown in Fig.~\ref{fig:C3_design}. 

\begin{figure*}[t!]
    \centering
    \makebox[\textwidth][c]{%
        \def\svgwidth{1.2\textwidth}
        \input{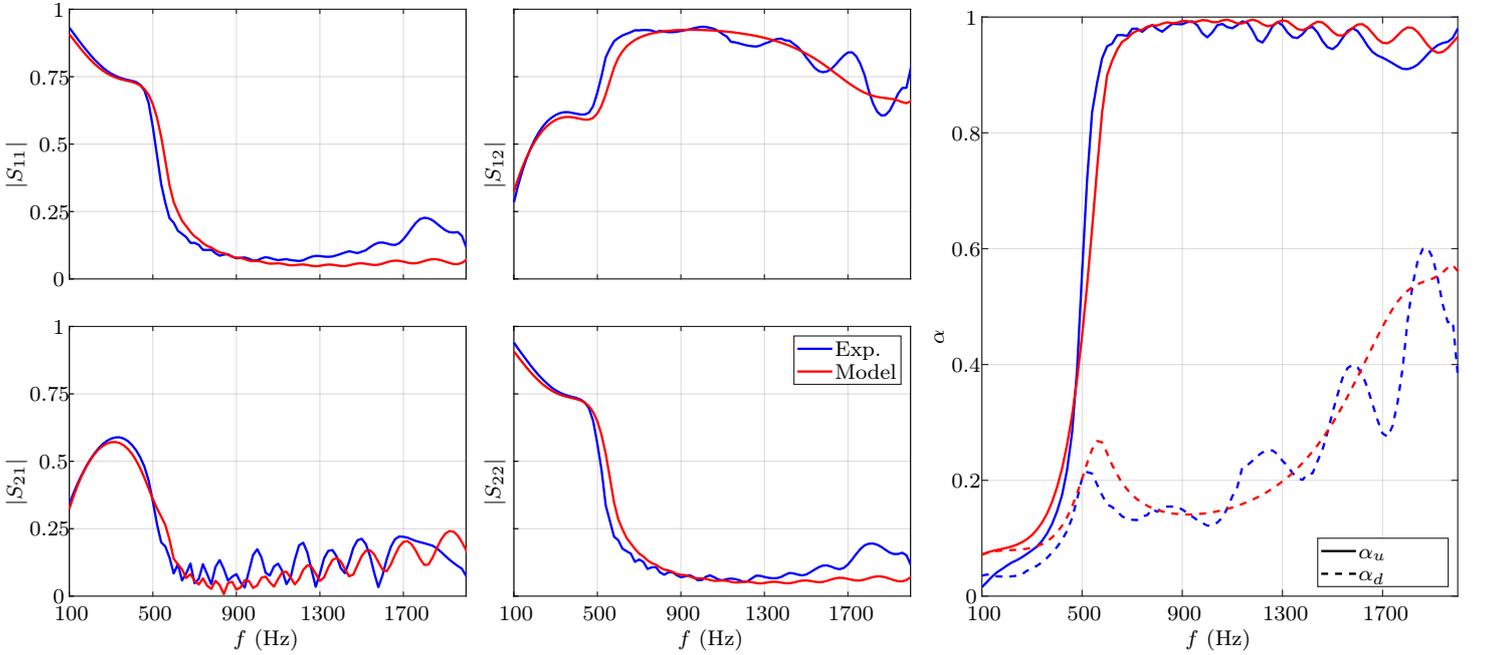}
    }
    \caption{Measured and modeled a) scattering matrix, b) dissipation coefficients of C3 design. $\alpha_u:$ upstream dissipation, $\alpha_d:$ downstream dissipation. }
    \label{fig:c3_exp_sim}
\end{figure*}

The comparison between the TMM model and the experiments of the scattering matrix and the dissipation coefficients for C3 are shown in Fig.~\ref{fig:c3_exp_sim}a) and b), respectively. As with the C1 and C2 designs, the model reproduces the measurements well across the relevant frequency range. The C3 design exhibits an effective cut-on frequency of \(520~\text{Hz}\), as shown in Fig.~\ref{fig:c3_exp_sim}b), where \(\alpha_u = 0.8\) at \(f = 520~\text{Hz}\). As in the C2 design, the dissipation coefficients remain above 0.95 up to \(2000~\text{Hz}\), underscoring the broadband performance of this relatively simple geometry.

\subsection{Reactive test-rig results}
\subsubsection{Upstream reflection coefficient measurements of the combustor setup}

The 3D-printed C3 ABH is fabricated from ABS, which cannot withstand the hot gases temperature generated by the flame. Accordingly, we place the ABH in the cold section of the test rig as shown in Fig.~\ref{fig:setup_reactive}. In the reactive configuration, as discussed earlier, the ABH is oriented so that the back cavity depth decreases along the flow direction, enhancing dissipation from the flame side. The measured pressure drop across the ABH at a given mass flow rate is identical to that without the ABH, indicating no pressure drop penalty from its implementation.

\begin{figure}[t!]
    \centering
    \def\svgwidth{0.7\textwidth}
    \input{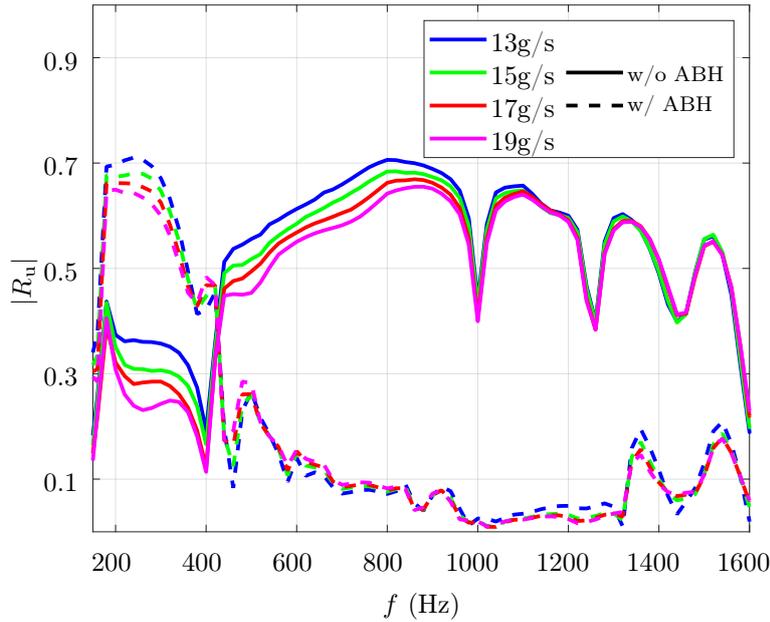}
    \caption{Comparison of the measured upstream reflection coefficient without (solid lines) and with (dashed lines) C3 ABH at different mass flow rates.}
    \label{fig:Rup_compare}
\end{figure}

The upstream reflection coefficient at the reference axial location, as indicated in \ref{fig:setup_reactive}, is measured under non-reactive conditions by forcing the system with loudspeakers downstream of the burner and placing four microphones next to the perforated ABH damper. Figure~\ref{fig:Rup_compare} presents the measured upstream reflection coefficients with and without the damper for different typical air mass flow rates through the combustor setup. In the absence of the ABH, the reflection coefficient of the inlet module, \(R_u\), is directly characterized. When the damper is installed, the effective reflection coefficient combines \(R_u\) with the scattering matrix elements of the ABH, as described in Eq.~\ref{eq:effective_ref}. The solid curves show that, at low frequencies, $R_u$ varies noticeably with air mass flow rate. However, above \(900~\text{Hz}\), these variations become much less pronounced. When the ABH is mounted, significant reductions in the reflection coefficient are observed for frequencies beyond 400~Hz.  However, below 400~Hz, the magnitude of the reflection coefficient is increased. This observation is consistent with the measured and modeled $|S_{12}|$ of the C3 design in Fig.~\ref{fig:c3_exp_sim}. For example, at $f=250$ Hz, $|S_{21}| = 0.6,~S_{11} = S_{22} = 0.8$, while $R_u \approx 0.3$. Hence, at this frequency, as indicated in Eq.~\ref{eq:effective_ref}, both $R_u$ and the scattering elements contribute to $R_{\mathrm{eff}}$, and because $0.75<|S_{11}| < 1$, $S_{12}$ is expected to dominate slightly. Above 500~Hz, $|S_{11}|$ and $|S_{22}|$ roll to zero quickly and therefore $R_u$ exerts a negligible influence on $R_{\mathrm{eff}}$ and $S_{21}$ dominates. When the damper is installed, \rx{$|R_{\mathrm{up}}|\leq0.1$} \rn{$|R_u|\leq0.1$} for $600~\mathrm{Hz}<f<1300~\mathrm{Hz}$, generating very low reflections.

\subsubsection{Self-excited dynamics with ABH}

\begin{figure}[t!]
    \centering
    \def\svgwidth{1\textwidth}
    \input{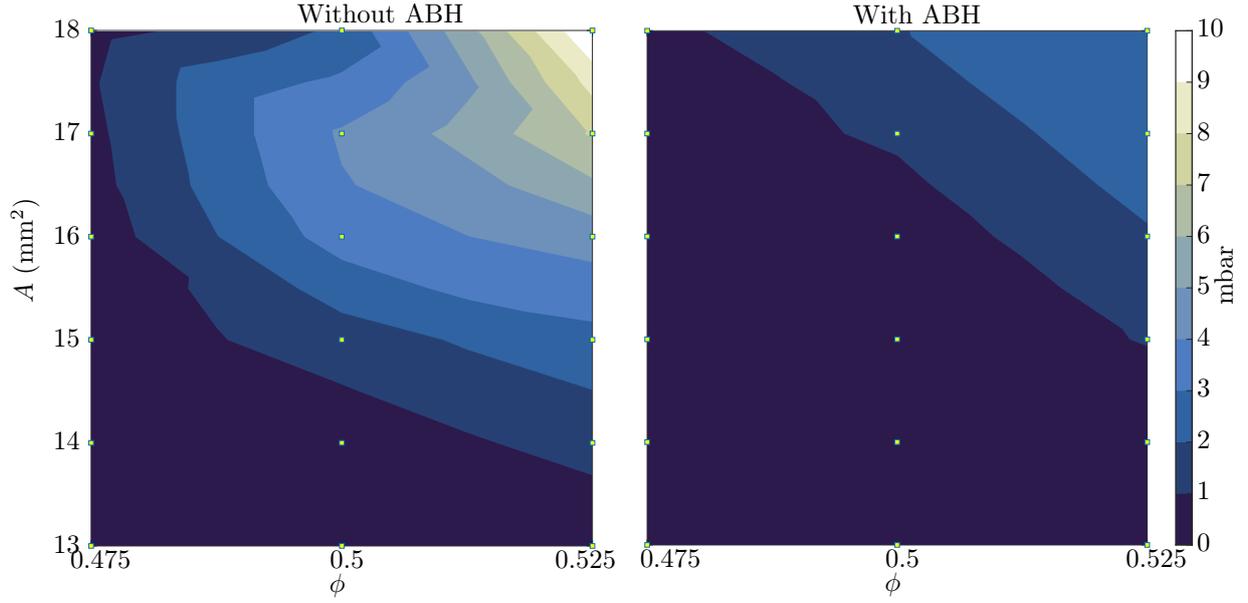}
    \caption{Pressure rms measured by the downstream microphone (see Fig.~\ref{fig:setup_reactive}) without ABH (left) and with ABH (right) at different equivalence ratios and outlet areas. The square markers are the measurement points.}
    \label{fig:rms_comp}
\end{figure}

The self-excited dynamics of the combustor at the three different equivalent ratios discussed in the previous section are investigated by recording the acoustic signals from the four microphones. Figure ~\ref{fig:rms_comp} shows the root mean squared (rms) values of the bandpass-filtered acoustic pressure signals measured by the downstream microphone.

Without the ABH, the combustor is unstable when the outlet area, $A>15~\mathrm{cm^2}$ and $\phi\geq0.5$. The highest pulsation is at $\phi = 0.525$ and $A = 18~\mathrm{cm^2}$ with $p_{rms} \approx 10$ mbar. For all operating conditions, the pulsation amplitude increases with the outlet area. 

\begin{figure}[t!]
    \centering
    \def\svgwidth{1\textwidth}
    \input{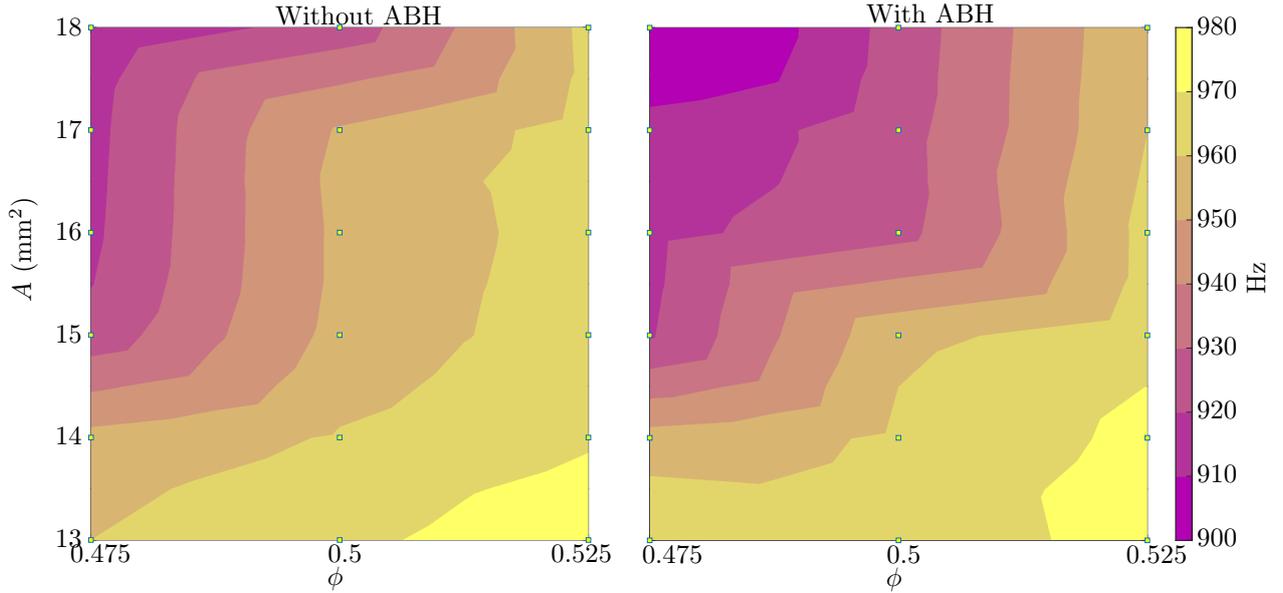}
    \caption{Peak frequency measured by downstream microphone (see Fig.~\ref{fig:setup_reactive}) without ABH (left) and with ABH (right). The square markers are the measurement points.}
    \label{fig:peak_comp}
\end{figure}

When the perforated ABH damper is installed, the pulsation decreases for all operating conditions. The stable operating point with $\phi=0.475$ remains stable for all outlet areas. For the case of $\phi = 0.5$, the combustor is fully stabilized at all areas. However, at $\phi = 0.525$, the combustor still exhibit thermoacoustic instabilities but with lower pulsation amplitudes of 2.5~mbar at $A = 18~\mathrm{cm^2}$, which corresponds to a reduction by a factor of four compared to the case without the damper. 
Figure.~\ref{fig:peak_comp} shows the peak frequency for the considered operating points and outlet areas. Across all cases, the peak frequency ranges from 900 to 980~Hz. A slight shift toward lower frequencies is observed when perforated ABH damper installed. 
\begin{figure*}[t!]
    \centering
    \def\svgwidth{1\textwidth}
    \input{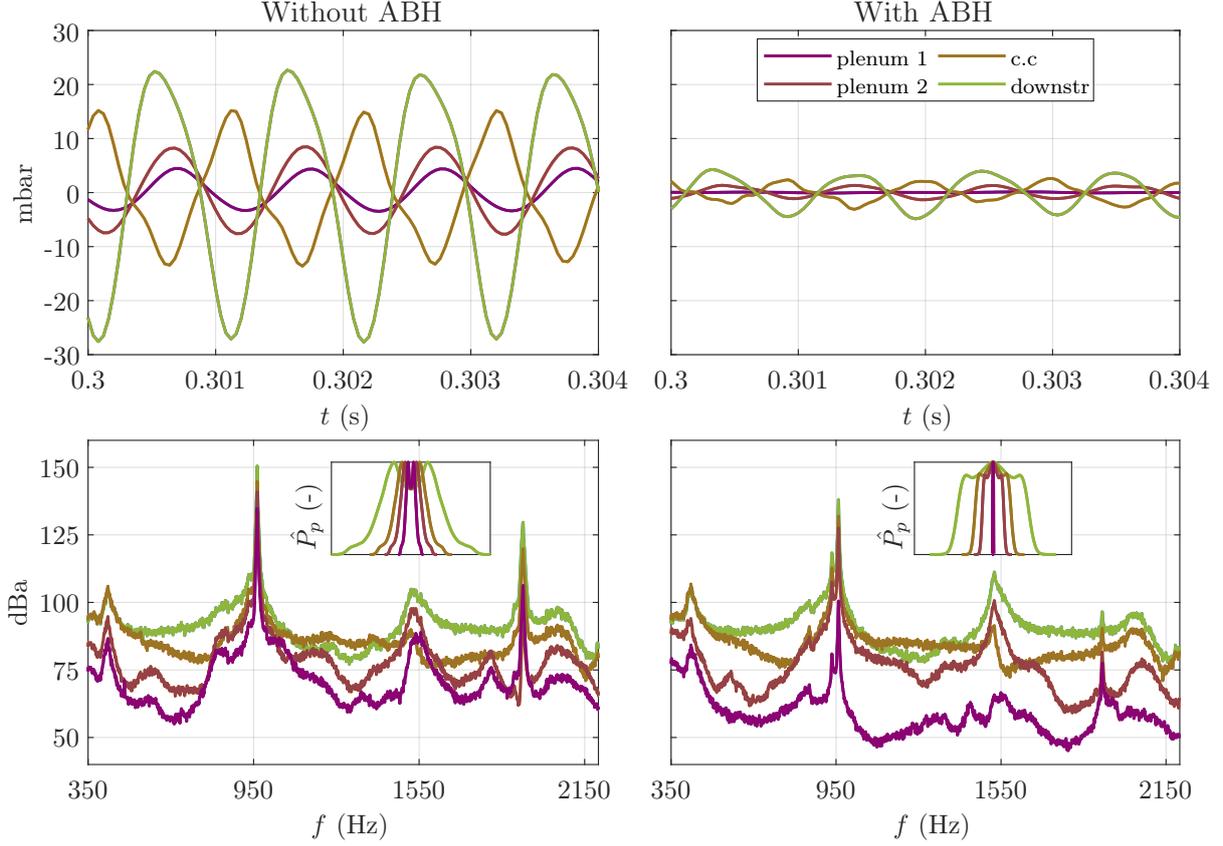}
    \caption{The time trace and acoustic power spectrum of the microphones (see Fig.~\ref{fig:setup_reactive}) without  (left) and with ABHs (right) at $\phi = 0.525,~A = 18~\mathrm{cm^2}$. The insets on the bottom plot show the distribution of the bandpass-filtered signals. Plenum~1 microphone measures significantly lower values compared to plenum~2 when ABHs are installed.}
    \label{fig:TT_PSD_comp}
\end{figure*}

\begin{figure*}[h]
    \centering
    \makebox[\textwidth][c]{%
        \def\svgwidth{1.1\textwidth}
        \input{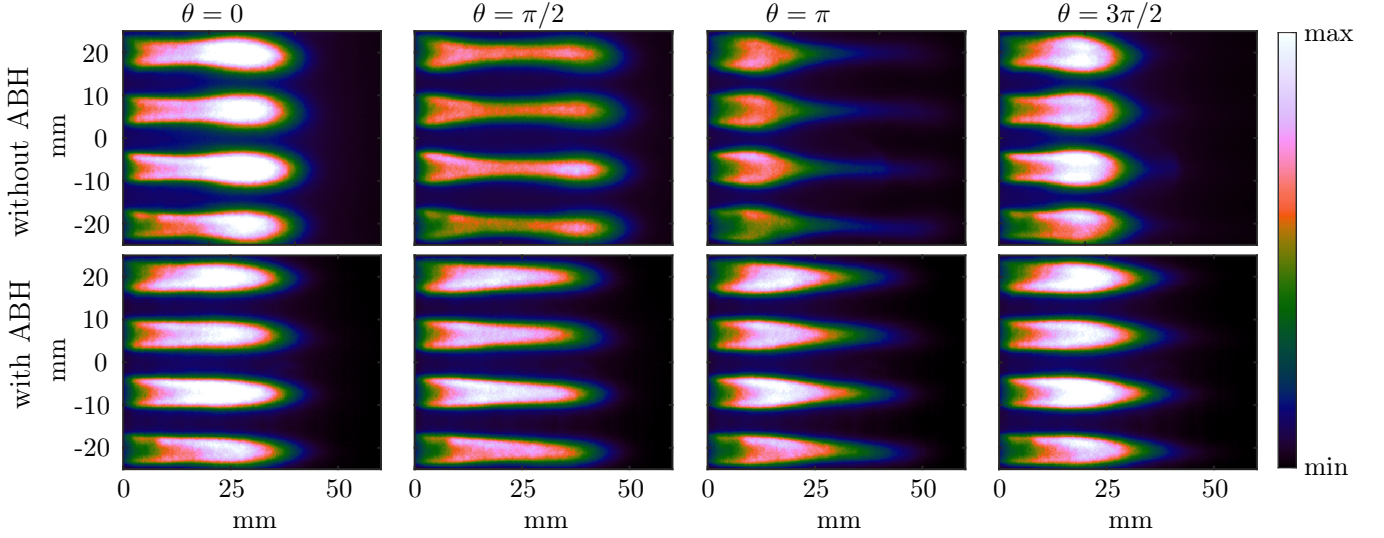}
    }
    \caption{Phase averaged flame $\mathrm{OH^*}$ chemiluminescence without (top row) and with (bottom row) perforated ABH damper. The phase angle is referenced to the integrated intensity signals.}
    \label{fig:phaseAvg}
\end{figure*}

Further discussion will be focused on the case of $\phi = 0.525$ which exhibits the highest pulsation amplitudes. Figure~\ref{fig:TT_PSD_comp} shows the time-trace of the acoustic signals and the power spectrum of all microphones at $\phi = 0.525,A = 18~\mathrm{cm^2}$. As seen, the highest amplitude is measured by the downstream microphone while, the two microphones in the plenum section read similar amplitudes. When the perforated ABH damper is installed, the overall pressure amplitudes decrease and the plenum~1 microphone detects very low values compared to the other microphones. From the corresponding power spectrum, it can be seen that the signal of plenum 1 and plenum 2 differs by about 25~dB for frequencies above 800~Hz. The insets of the bottom plots show the normalized probability density function (pdf) of the pressure signals, $\hat{P}_p$. Without the ABH, the bimodal distribution is prominent, which is a strong feature of an unstable system. Whereas, without the damper, trimodal distributions are observed, this indicates that the system is bistable. Further analysis can be performed to study this interesting behavior, however, this lies outside the scope of the current study. Figure.~\ref{fig:phaseAvg} shows the corresponding phase averaged flame $\mathrm{OH^*}$ chemiluminescence at four different phase angles. As can be seen, without the perforated ABH, the flame oscillates strongly in the longitudinal direction, and the intensity fluctuates significantly between the phase angles. When the damper is installed, because the system is also unstable, the flame still oscillates albeit with significantly lower amplitudes.

\begin{figure}[h]
    \centering

        \makebox[\textwidth][c]{%
    \def\svgwidth{1.1\textwidth}
    \input{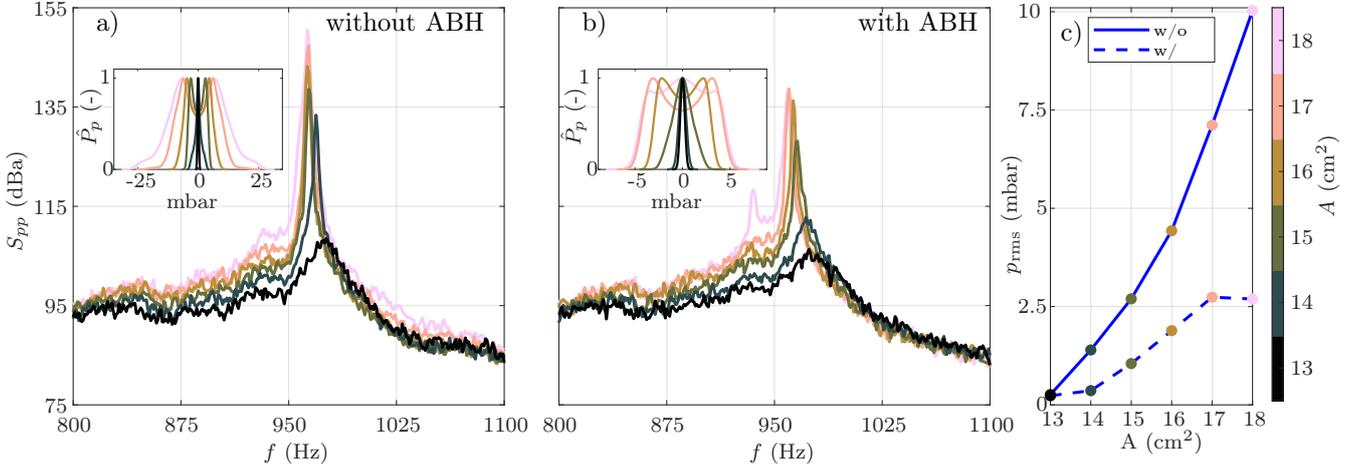}
    }

    \caption{Power spectrum of the acoustic pressure measured by the downstream microphone at $\phi = 0.525$ for different outlet areas: (a) without and (b) with the perforated ABH damper. The insets show the scaled probability density function of the bandpass-filtered acoustic signal. (c) Root-mean-square pressure as a function of outlet area for cases without (solid line) and with (dashed-line) the ABH damper.}
    \label{fig:OP14_PSD_compare}
\end{figure}

Figure~\ref{fig:OP14_PSD_compare} presents the power spectra, the normalized pdf of the band-pass-filtered acoustic pressure, $\hat{P}p$, and the root-mean-square pressure, $p_{\mathrm{rms}}$, at $\phi = 0.525$ for different outlet areas, measured by the downstream microphone for configurations with and without the ABH. As shown in Fig.~\ref{fig:OP14_PSD_compare}(a), in the absence of the ABH, the system remains stable at $A = 13~\mathrm{cm^2}$, as indicated by the unimodal distribution of $\hat{P}_p$. At $A = 14~\mathrm{cm^2}$, the system becomes marginally stable; the pressure pdf broadens noticeably, and the acoustic power at the peak frequency increases sharply from approximately 110~dBa to nearly 135~dBa. At $A = 15~\mathrm{cm^2}$, the system transitions to thermoacoustic instability, as evidenced by the emergence of a bimodal distribution in $\hat{P}_p$. The pressure rms increases monotonically with outlet area as shown in Fig.~\ref{fig:OP14_PSD_compare}c. For the configuration with the ABH, Fig.~\ref{fig:OP14_PSD_compare}(b) shows that at $A = 15~\mathrm{cm^2}$ the distribution is no longer bimodal, but instead exhibits a broadened unimodal shape, while $p_{\mathrm{rms}}$ is approximately 1 mbar, corresponding to less than half the value observed without the ABH (see Fig.~\ref{fig:OP14_PSD_compare}c). The pressure rms increases further with outlet area until reaching a plateau at $A = 17~\mathrm{cm^2}$. At the largest outlet area, $\hat{P}_p$ is no longer bimodal but instead exhibits a trimodal distribution. Although the ABH does not fully stabilize the system, it yields a substantial reduction in pressure amplitude, highlighting its effectiveness as a passive damper. \rev{It is acknowledged that, for the present configuration, stabilization of the investigated operating points can also be achieved by adjusting the tunable outlet boundary conditions in the absence of the perforated ABH damper. However, such tunable boundaries are specific to the experimental setup, require substantial and costly engineering effort, and are not readily transferable to practical industrial combustor applications.}

It is expected that placing the ABH on the walls of the cage modules downstream of the flame will be more effective, as typically the pressure anti-node of the thermoacoustic mode occurs inside the combustion chamber. However, there are additional challenges that need to be overcome. First, the ABH have to be made out of metal. Second, the ABH blocks have to be water cooled to avoid overheating. Third, the cavities have to be purged with air so that the resonance frequencies remain at the desired points. Purging the air cavities with a bias flow will indeed increase the resistance of the \rx{perforator}\rn{perforated plate}s as shown by studies on combustor liner damper \cite{LAHIRI2017,Bellucci2002,Bellucci2004,LUONG2005,Miniero2023}. Furthermore, the interaction of the grazing flow and the bias flow will also change the effective impedance of \rx{perforator}\rn{perforated plate}s, depending on the Mach number of each stream \cite{Kim2018,Elnady2003}. These coupled flow–acoustic effects \rn{and the underlying nonlinearities \cite{Miniero2023}} motivate a dedicated parametric assessment, which will be addressed in future work. Notwithstanding these challenges, the present results demonstrate strong potential for perforated ABH-based damping and provide a sound baseline for selecting geometric parameters for the metallic version of the damper.
\section{Conclusion}

In this study, a reduced-order model based on the transfer matrix method (TMM) was developed to predict the acoustic scattering behavior of perforated acoustic black hole (ABH) dampers with graded geometric parameters. The model was validated against scattering-matrix measurements obtained from additively manufactured ABH prototypes tested in a non-reactive acoustic rig, showing good agreement over the frequency range of interest. Leveraging this validated framework, a damper design was optimized to maximize acoustic dissipation over a broad frequency band, such that the \rn{frequency of the } dominant thermoacoustic instability  of a technically premixed hydrogen combustor was encompassed within the effective dissipation bandwidth. The optimized ABH damper was subsequently installed in the plenum section of the combustor test rig, where reflection coefficient measurements revealed a substantial reduction of the upstream reflection coefficient for frequencies above 500~Hz.

The thermoacoustic performance of the ABH damper was further assessed under reactive conditions by measuring acoustic pressure oscillations for hydrogen flames over a range of equivalence ratios and outlet boundary areas. For all operating conditions investigated, the presence of the ABH damper led to a marked attenuation of acoustic pulsation amplitudes compared to configurations without the damper. Complete stabilization was achieved at $\phi = 0.5$, while at $\phi = 0.525$ and outlet area, $A \geq 16~\mathrm{cm^2}$, the system remained unstable but exhibited a reduction in pulsation amplitudes by approximately a factor of four. These results demonstrate that the perforated ABH damper provides robust and broadband thermoacoustic damping in a hydrogen-fueled combustor, even under conditions where complete stabilization is not achieved.

Future work will focus on extending the present concept toward more practical implementations by manufacturing the damper from metallic materials and installing it closer to the combustion chamber, where acoustic pressure antinodes are typically located. Such configurations are expected to enhance coupling between the damper and the unstable acoustic modes, thereby further improving stabilization performance and facilitating the transfer of the proposed concept to industrially relevant combustor architectures.


\section*{Author contributions}

B.D. conceived the research project \rx{. N.N Supervised the project. B.D.}\rn{ and }led the experimental and modeling investigations. B.D and K.C performed the experiments and developed the model. All authors discussed the results. B.D. wrote the  paper. \rn{N.N reviewed and edited the paper, and supervised the project.} The final version of the manuscript has been \rx{edited and} approved by all the authors.

\bibliographystyle{elsarticle-num}

\bibliography{ref}

@article{Mironov,
	author = {Mironov, M. A. and Pislyakov, V. V.},
	journal = {Acoustical Physics},
	number = {3},
	pages = {347--352},
	title = {One-dimensional acoustic waves in retarding structures with propagation velocity tending to zero},
	volume = {48},
	year = {2002}}

@ARTICLE{Martin2025,
	author = {Martin, Richard and Pandey, Khushboo and Schuermans, Bruno and Noiray, Nicolas},
	title = {Orifice whistling suppression with slow sound},
	year = {2025},
	journal = {Journal of Fluid Mechanics},
	volume = {1018},
	doi = {10.1017/jfm.2025.10569}}

@ARTICLE{Stoychev2024a,
	author = {Stoychev, Alexander K. and Noiray, Nicolas},
	title = {Nonlinear dynamics of an acoustically compact orifice},
	year = {2024},
	journal = {Journal of Sound and Vibration},
	volume = {593},
	doi = {10.1016/j.jsv.2024.118660}}

@ARTICLE{Stoychev2024b,
	author = {Stoychev, Alexander K. and Pedergnana, Tiemo and Noiray, Nicolas},
	title = {Nonlinear acoustics of an aperture under grazing flow},
	year = {2024},
	journal = {Proceedings of the Royal Society A: Mathematical, Physical and Engineering Sciences},
	volume = {480},
	number = {2283},
	doi = {10.1098/rspa.2023.0718}}

@ARTICLE{Miniero2023,
	author = {Miniero, Luigi and Mensah, Georg A. and Bourquard, Claire and Noiray, Nicolas},
	title = {Failure of thermoacoustic instability control due to periodic hot gas ingestion in Helmholtz dampers},
	year = {2023},
	journal = {Journal of Sound and Vibration},
	volume = {548},
	doi = {10.1016/j.jsv.2022.117544}}

@article{PELAT2020115316,
	author = {Adrien Pelat and Fran{\c c}ois Gautier and Stephen C. Conlon and Fabio Semperlotti},
	issn = {0022-460X},
	journal = {Journal of Sound and Vibration},
	pages = {115316},
	title = {The acoustic black hole: A review of theory and applications},
	volume = {476},
	year = {2020},}

@article{ElOuahabi,
    author = {Azbaid El Ouahabi, Abdelhalim and Krylov, Victor and O'Boy, Dan},
    year = {2015},
    month = {07},
    pages = {},
    title = {Experimental investigation of the acoustic black hole for sound absorption in air},
    journal = {22nd International Congress on Sound and Vibration},
}

@article{Guash,
	author = {Oriol Guasch and Marc Arnela and Patricia S{\'a}nchez-Mart{\'\i}n},
	journal = {Journal of Sound and Vibration},
	pages = {65-79},
	title = {Transfer matrices to characterize linear and quadratic acoustic black holes in duct terminations},
	volume = {395},
	year = {2017}}

@article{Li2024,
	author = {Sihui Li and Jiajun Xia and Xiang Yu and Xiaoqi Zhang and Li Cheng},
	journal = {Journal of Sound and Vibration},
	pages = {117781},
	title = {A sonic black hole structure with perforated boundary for slow wave generation},
	volume = {559},
	year = {2023},}

@article{BaranekIngard,
	author = {Noureddine Atalla and Franck Sgard},
	journal = {Journal of Sound and Vibration},
	number = {1},
	pages = {195-208},
	title = {Modeling of perforated plates and screens using rigid frame porous models},
	volume = {303},
	year = {2007},}

@article{Jimenez2017,
  author       = {Jiménez, Noé and Romero-García, Vicent and Pagneux, Vincent and Groby, Jean-Philippe},
  title        = {Rainbow-trapping absorbers: Broadband, perfect and asymmetric sound absorption by subwavelength panels for transmission problems},
  journal      = {Scientific Reports},
  year         = {2017},
  volume       = {7},
  number       = {1},
  pages        = {13595},
}

@article{GUASCH2017,
title = {Transfer matrices to characterize linear and quadratic acoustic black holes in duct terminations},
journal = {Journal of Sound and Vibration},
volume = {395},
pages = {65-79},
year = {2017},
issn = {0022-460X},
author = {Oriol Guasch and Marc Arnela and Patricia Sánchez-Martín},
}

@article{Bravo2023,
  title = {Broadband sound attenuation and absorption by duct silencers based on the acoustic black hole effect: Simulations and experiments},
  author = {Bravo, Teresa and Maury, C{\'e}dric},
  journal = {Journal of Sound and Vibration},
  volume = {561},
  pages = {117925},
  year = {2023},
  publisher = {Elsevier}
}

@article{Bravo2024,
  title = {Converging rainbow trapping silencers for broadband sound dissipation in a low-speed ducted flow},
  author = {Bravo, Teresa and Maury, C{\'e}dric},
  journal = {Journal of Sound and Vibration},
  volume = {589},
  pages = {118524},
  year = {2024},
  publisher = {Elsevier},
}

@article{Maury2025,
  title = {Micro-perforated rainbow-trapping silencers with broadband sound dissipation and reduced drag under low-speed grazing flow},
  author = {Maury, C{\'e}dric and Bravo, Teresa and Ali, Fawad},
  journal = {Journal of Sound and Vibration},
  volume = {616},
  pages = {119228},
  year = {2025},
  publisher = {Elsevier},
}

@article{DHARMAPUTRA2023,
title = {Thermoacoustic stabilization of a sequential combustor with ultra-low-power nanosecond repetitively pulsed discharges},
journal = {Combust. Flame.},
volume = {258},
pages = {113101},
year = {2023},
issn = {0010-2180},
author = {Bayu Dharmaputra and Sergey Shcherbanev and Bruno Schuermans and Nicolas Noiray},
}

@article{DHARMAPUTRA2024,
title = {BOATS: Bayesian Optimization for Active control of ThermoacousticS},
journal = {J. Sound. Vib},
volume = {582},
pages = {118415},
year = {2024},
author = {Bayu Dharmaputra and Pit Reckinger and Bruno Schuermans and Nicolas Noiray},
}

@article{Pandalai1998,
author = {Raghavan Pandalai and Hukam Mongia},
title = {Combustion instability characteristics of industrial engine dry low emission combustion systems},
Journal = {AIAA Meeting Paper},
pages = {3379},
year = {1998}
}

@article{Bellucci2004,
author = {Bellucci, V. and Flohr, P. and Paschereit, C. O. and Magni, F.},
issn = {07424795},
journal = {J. Eng. Gas. Turb. Power},
pages = {271--275},
title = {{On the use of Helmholtz resonators for damping acoustic pulsations in industrial gas turbines}},
volume = {126},
year = {2004}
}

@article{Moon2024,
  title = {Transfer functions of lean fully- and technically-premixed jet-stabilized turbulent hydrogen flames},
  author = {Moon, Kihun and Martin, Richard and Schuermans, Bruno and Noiray, Nicolas},
  journal = {Proceedings of the Combustion Institute},
  volume = {40},
  pages = {105256},
  year = {2024},
  publisher = {Elsevier}
}

@article{Eirik2023,
    author = {Æsøy, Eirik and Indlekofer, Thomas and Bothien, Mirko R. and Dawson, James R.},
    title = {The Effect of Hydrogen on Nonlinear Flame Saturation},
    journal = {Journal of Engineering for Gas Turbines and Power},
    volume = {145},
    number = {11},
    pages = {111019},
    year = {2023},
    month = {09},
}

@article{FAUREBEAULIEU2024,
title = {Measuring acoustic transfer matrices of high-pressure hydrogen/air flames for aircraft propulsion},
journal = {Combustion and Flame},
volume = {270},
pages = {113776},
year = {2024},
issn = {0010-2180},
author = {Abel Faure-Beaulieu and Bayu Dharmaputra and Bruno Schuermans and Guoqing Wang and Stephan Caruso and Maximilian Zahn and Nicolas Noiray},
}

@article{SERRA2023,
title = {Optimization of the profile and distribution of absorption material in sonic black holes},
journal = {Mechanical Systems and Signal Processing},
volume = {202},
pages = {110707},
year = {2023},
issn = {0888-3270},
author = {Gerard Serra and Oriol Guasch and Marc Arnela and David Miralles},
}

@article{BROOKE2020,
title = {Acoustic metamaterial for low frequency sound absorption in linear and nonlinear regimes},
journal = {Journal of Sound and Vibration},
volume = {485},
pages = {115585},
year = {2020},
issn = {0022-460X},
author = {Daniel C. Brooke and Olga Umnova and Philippe Leclaire and Thomas Dupont},
}

@ARTICLE{Johnson1987,
	author = {Johnson, David Linton and Koplik, Joel and Dashen, Roger},
	title = {Theory of dynamic permeability and tortuosity in fluid saturated porous media},
	year = {1987},
	journal = {Journal of Fluid Mechanics},
	volume = {176},
	pages = {379 – 402},
}

@ARTICLE{Champoux1991,
	author = {Champoux, Yvan and Allard, Jean-F.},
	title = {Dynamic tortuosity and bulk modulus in air-saturated porous media},
	year = {1991},
	journal = {Journal of Applied Physics},
	volume = {70},
	number = {4},
	pages = {1975 – 1979},
}

@ARTICLE{Lafarge1997,
	author = {Lafarge, Denis and Lemarinier, Pavel and Allard, Jean F. and Tarnow, Viggo},
	title = {Dynamic compressibility of air in porous structures at audible frequencies},
	year = {1997},
	journal = {Journal of the Acoustical Society of America},
	volume = {102},
	number = {4},
	pages = {1995 – 2006},
}

@article{ZHU2025,
title = {Acoustic properties analysis of ABH complex structures with micro-perforated boundaries},
journal = {Applied Acoustics},
volume = {237},
pages = {110752},
year = {2025},
issn = {0003-682X},
author = {Hanya Zhu and Xiao Liang and Nansha Gao and Liang Shi},
}

@article{li_slow_2025,
	title = {Slow waves in ducts with external SBH insertion and perforated boundaries},
	volume = {236},
	issn = {0003682X},
	journal = {Applied Acoustics},
	author = {Li, Sihui and Yu, Xiang and Cheng, Li},
	month = jun,
	year = {2025},
	pages = {110754},
}

@inproceedings{Maury2025Perforated,
  author    = {Maury, C{\'e}dric and Bravo, Teresa and Mazzoni, Daniel},
  editor    = {Doolan, Con and Moreau, Danielle and Wills, Angus},
  title     = {Modelling and Characterization of Micro-Porous Resonating Liners under a Low-Speed Flow},
  booktitle = {Flow Induced Noise and Vibration Issues and Aspects IV},
  year      = {2025},
  publisher = {Springer Nature Switzerland},
  pages     = {177--203}
}

@article{Paschereit2002,
    author = {Paschereit, C. O. and Schuermans, B. and Polifke, W. and Mattson, O.},
    title = "{Measurement of Transfer Matrices and Source Terms of Premixed Flames }",
    journal = {J. Eng. Gas Turbine. Power},
    volume = {124},
    number = {2},
    pages = {239-247},
    year = {2002},
    month = {03},
}

@inproceedings{Schuermans2004,
    author = {Schuermans, Bruno and Bellucci, Valter and Guethe, Felix and Meili, Franc¸ois and Flohr, Peter and Paschereit, Christian Oliver},
    title = {A Detailed Analysis of Thermoacoustic Interaction Mechanisms in a Turbulent Premixed Flame},
    booktitle = {Proceedings of the ASME Turbo Expo 2004},
    pages = {539-551},
    year = {2004},
    month = {06},
}

@article{Maa1998,
    author = {Maa, Dah-You},
    title = {Potential of microperforated panel absorber},
    journal = {The Journal of the Acoustical Society of America},
    volume = {104},
    number = {5},
    pages = {2861-2866},
    year = {1998},
    month = {11},
}

@article{temiz_2016,
	title = {Non-linear acoustic transfer impedance of micro-perforated plates with circular orifices},
	volume = {366},
	urldate = {2026-01-13},
	journal = {Journal of Sound and Vibration},
	author = {Temiz, Muttalip Aşkın and Tournadre, Jonathan and Arteaga, Ines Lopez and Hirschberg, Avraham},
	year = {2016},
	pages = {418--428},
}

@article{HUMBERT2025,
title = {Nonlinear acoustic resistance of perforated plates with two high-amplitude harmonic excitations and a steady bias flow},
journal = {Journal of Sound and Vibration},
volume = {618},
pages = {119315},
year = {2025},
author = {Sylvain C. Humbert},
}

@article{LAHIRI2017,
title = {A review of bias flow liners for acoustic damping in gas turbine combustors},
journal = {Journal of Sound and Vibration},
volume = {400},
pages = {564-605},
year = {2017},
author = {C. Lahiri and F. Bake},
}

@article{MELLING1973,
title = {The acoustic impendance of perforates at medium and high sound pressure levels},
journal = {Journal of Sound and Vibration},
volume = {29},
number = {1},
pages = {1-65},
year = {1973},
issn = {0022-460X},
author = {T.H. Melling},
}

@article{BLONDE2025,
title = {Tunable passive control of thermoacoustic instabilities based on a variable geometry combustor outlet nozzle},
journal = {Aerospace Science and Technology},
volume = {158},
pages = {109932},
year = {2025},
issn = {1270-9638},
author = {Audrey Blondé and Bruno Schuermans and Bayu Dharmaputra and Nicolas Noiray}
}

@inproceedings{Bellucci2002,
  author       = {Bellucci, Valter and Paschereit, Christian Oliver and Flohr, Peter},
  title        = {Impedance of {P}erforated {S}creens with {B}ias {F}low},
  booktitle    = {8th {AIAA}/{CEAS} Aeroacoustics Conference \& Exhibit},
  address      = {Breckenridge, Colorado, USA},
  month        = jun,
  year         = {2002},
  number       = {AIAA 2002-2437},
  pages={1--9}
}

@article{LUONG2005,
title = {On the Rayleigh conductivity of a bias-flow aperture},
journal = {Journal of Fluids and Structures},
volume = {21},
number = {8},
pages = {769-778},
year = {2005},
author = {T. Luong and M.S. Howe and R.S. McGowan},
}

@article{Kim2018,
    author = {Kim, Daesik and Jung, Seungchai and Park, Heeho},
    title = {Design of Acoustic Liner in Small Gas Turbine Combustor Using One-Dimensional Impedance Models},
    journal = {Journal of Engineering for Gas Turbines and Power},
    volume = {140},
    number = {12},
    pages = {121505},
    year = {2018},
    month = {08},
	}

@inproceedings{Elnady2003,
author = {Tamer Elnady and Hans Boden},
title = {On Semi-Empirical Liner Impedance Modeling with Grazing Flow},
booktitle = {9th AIAA/CEAS Aeroacoustics Conference and Exhibit},
year = {2003},
pages={1--11}
}

@article{HOEIJMAKERS2014,
title = {Intrinsic instability of flame–acoustic coupling},
journal = {Combustion and Flame},
volume = {161},
number = {11},
pages = {2860-2867},
year = {2014},
issn = {0010-2180},
author = {Maarten Hoeijmakers and Viktor Kornilov and Ines {Lopez Arteaga} and Philip {de Goey} and Henk Nijmeijer},
}

@article{EMMERT2015,
title = {Intrinsic thermoacoustic instability of premixed flames},
journal = {Combustion and Flame},
volume = {162},
number = {1},
pages = {75-85},
year = {2015},
issn = {0010-2180},
author = {Thomas Emmert and Sebastian Bomberg and Wolfgang Polifke},
}

@article{EMMERT2017,
title = {Acoustic and intrinsic thermoacoustic modes of a premixed combustor},
journal = {Proceedings of the Combustion Institute},
volume = {36},
number = {3},
pages = {3835-3842},
year = {2017},
author = {T. Emmert and S. Bomberg and S. Jaensch and W. Polifke},
}

@article{SILVA2023,
title = {Intrinsic thermoacoustic instabilities},
journal = {Progress in Energy and Combustion Science},
volume = {95},
pages = {101065},
year = {2023},
author = {Camilo F. Silva},
}

@article{DHARMAPUTRA2024_PROCI,
title = {Plasma assisted thermoacoustic stabilization of a transiently operated sequential combustor at high pressure},
journal = {Proceedings of the Combustion Institute},
volume = {40},
number = {1},
pages = {105518},
year = {2024},
author = {Bayu Dharmaputra and Sergey Shcherbanev and Nicolas Noiray},
}

@article{Bothien2013,
    author = {Bothien, Mirko R. and Noiray, Nicolas and Schuermans, Bruno},
    title = {A Novel Damping Device for Broadband Attenuation of Low-Frequency Combustion Pulsations in Gas Turbines},
    journal = {Journal of Engineering for Gas Turbines and Power},
    volume = {136},
    number = {4},
    pages = {041504},
    year = {2013},
    month = {12},
}

@article{Bellucci2004_HHD,
    author = {Bellucci, V.  and Flohr, P.  and Paschereit, C. O.  and Magni, F. },
    title = {{On the Use of Helmholtz Resonators for Damping Acoustic Pulsations in Industrial Gas Turbines} },
    journal = {Journal of Engineering for Gas Turbines and Power},
    volume = {126},
    number = {2},
    pages = {271-275},
    year = {2004},
    month = {06},
}

@article{Bellucci2005,
    author = {Bellucci, Valter and Schuermans, Bruno and Nowak, Dariusz and Flohr, Peter and Paschereit, Christian Oliver},
    title = {Thermoacoustic Modeling of a Gas Turbine Combustor Equipped With Acoustic Dampers },
    journal = {Journal of Turbomachinery},
    volume = {127},
    number = {2},
    pages = {372-379},
    year = {2005},
    month = {05},
}

@article{schnell2009,
    author = {Schnell, Alexander and Noiray, Nicolas and Reinert, Felix and Lauffer, Diane and Schuermans, Bruno},
    title = {Combustion device of a gas turbine },
   journal = {EP2295864B1},
       year = {2009}       
      }

@ARTICLE{Noiray20122753,
	author = {Noiray, Nicolas and Schuermans, Bruno},
	title = {Theoretical and experimental investigations on damper performance for suppression of thermoacoustic oscillations},
	year = {2012},
	journal = {Journal of Sound and Vibration},
	volume = {331},
	number = {12},
	pages = {2753 – 2763},
	doi = {10.1016/j.jsv.2012.02.005},
	}

@article{Noble2021,
    author = {Noble, David and Wu, David and Emerson, Benjamin and Sheppard, Scott and Lieuwen, Tim and Angello, Leonard},
    title = {Assessment of Current Capabilities and Near-Term Availability of Hydrogen-Fired Gas Turbines Considering a Low-Carbon Future},
    journal = {Journal of Engineering for Gas Turbines and Power},
    volume = {143},
    number = {4},
    pages = {041002},
    year = {2021},
    month = {02},
}

@article{OBERG2022,
title = {The value of flexible fuel mixing in hydrogen-fueled gas turbines – A techno-economic study},
journal = {International Journal of Hydrogen Energy},
volume = {47},
number = {74},
pages = {31684-31702},
year = {2022},
author = {Simon Öberg and Mikael Odenberger and Filip Johnsson},
}

@article{DHARMAPUTRA2025,
title = {Flame transfer function measurement of a sequential combustor fuelled with natural gas and hydrogen},
journal = {Combustion and Flame},
volume = {274},
pages = {113972},
year = {2025},
author = {Bayu Dharmaputra and Pushkin Nagpure and Matteo Impagnatiello and Nicolas Noiray},
}

@inproceedings{Ciani2021HydrogenBlending,
  author    = {Ciani, A. and Tay-Wo-Chong, L. and Amato, A. and Bertolotto, E. and Spataro, G.},
  title     = {Hydrogen Blending Into Ansaldo Energia AE94.3A Gas Turbine: High Pressure Tests, Field Experience and Modelling Considerations},
  booktitle = {Proceedings of the ASME Turbo Expo 2021},
  year      = {2021},
  address   = {Virtual, Online},
  month     = {June},
  pages     = {V03AT04A009},
  publisher = {ASME},
  doi       = {10.1115/GT2021-58650}
}

@inproceedings{Aoki2024HydrogenMicromix,
  author    = {Aoki, S. and Uto, T. and Takahashi, N. and Okada, K. and Kroniger, D. and Kamiya, H. and Yamaguchi, M. and Ishimura, Y. and Wirsum, M. and Funke, H. H. and Kusterer, K.},
  title     = {Development of Hydrogen and Micromix Combustor for Small and Medium Size Gas Turbine of Kawasaki},
  booktitle = {Proceedings of the ASME Turbo Expo 2024},
  year      = {2024},
  address   = {London, United Kingdom},
  month     = {June},
  pages     = {V002T03A003},
  publisher = {ASME},
  doi       = {10.1115/GT2024-121073}
}

@inproceedings{Clemen2024HydrogenAeroGT,
  author    = {Clemen, C. and Ravikanti, M. and La Bianca, N. and Eggels, R. and Wurm, B. and Young, K.},
  title     = {Considerations for Hydrogen Fueled Aerospace Gas Turbine Combustion Sub-System Design},
  booktitle = {Proceedings of the ASME Turbo Expo 2024},
  year      = {2024},
  address   = {London, United Kingdom},
  month     = {June},
  pages     = {V03AT04A016},
  publisher = {ASME},
}

@ARTICLE{Bourquard2019288,
	author = {Bourquard, C. and Noiray, N.},
	title = {Stabilization of acoustic modes using Helmholtz and Quarter-Wave resonators tuned at exceptional points},
	year = {2019},
	journal = {Journal of Sound and Vibration},
	volume = {445},
	pages = {288 – 307},
	doi = {10.1016/j.jsv.2018.12.011},
}

@article{AGUILAR2022,
title = {The influence of hydrogen on the stability of a perfectly premixed combustor},
journal = {Combustion and Flame},
volume = {245},
pages = {112323},
year = {2022},
issn = {0010-2180},
author = {José G. Aguilar and Eirik Æsøy and James R. Dawson},
}

@article{LEE2020,
title = {Combustion dynamics of lean fully-premixed hydrogen-air flames in a mesoscale multinozzle array},
journal = {Combustion and Flame},
volume = {218},
pages = {234-246},
year = {2020},
author = {Taesong Lee and Kyu Tae Kim},
}

@article{SCARPATO2012,
title = {Modeling the damping properties of perforated screens traversed by a bias flow and backed by a cavity at low Strouhal number},
journal = {Journal of Sound and Vibration},
volume = {331},
number = {2},
pages = {276-290},
year = {2012},
author = {A. Scarpato and N. Tran and S. Ducruix and T. Schuller},
}

@article{ARAVIND2024,
title = {Understanding the coupling between nanosecond repetitively pulsed discharges and the thermoacoustic behavior of a swirl flame at 2 bar},
journal = {Proceedings of the Combustion Institute},
volume = {40},
number = {1},
pages = {105211},
year = {2024},
author = {B Aravind and Liang Yu and Deanna A. Lacoste},
}

@article{Mi2021,
    author = {Mi, Yongzhen and Zhai, Wei and Cheng, Li and Xi, Chenyang and Yu, Xiang},
    title = {Wave trapping by acoustic black hole: Simultaneous reduction of sound reflection and transmission},
    journal = {Applied Physics Letters},
    volume = {118},
    number = {11},
    pages = {114101},
    year = {2021},
    month = {03},
}

@article{Mironov1988,
  author  = {Mironov, M. A.},
  title   = {Propagation of a flexural wave in a plate whose thickness decreases smoothly to zero in a finite interval},
  journal = {Sov. Phys. Acoust.},
  volume  = {34},
  number  = {3},
  year    = {1988},
  pages   = {318--319}
}

@inproceedings{Krylov2007_ABH,
  author    = {Krylov, V. V.},
  title     = {Propagation of plate bending waves in the vicinity of one- and two-dimensional acoustic black holes},
  booktitle = {ECCOMAS Thematic Conference on Computational Methods},
  year      = {2007},
  address   = {Rethymno, Crete, Greece},
  pages   = {1--12}
}

@article{HRUSKA2024,
title = {Complex frequency analysis and source of losses in rectangular sonic black holes},
journal = {Journal of Sound and Vibration},
volume = {571},
pages = {118107},
year = {2024},
author = {Viktor Hruška and Jean-Philippe Groby and Michal Bednařík},
}

@article{Petrover2024,
    author = {Petrover, Kayla and Baz, A.},
    title = {Acoustic black hole with functionally graded perforated rings},
    journal = {Journal of Applied Physics},
    volume = {135},
    number = {23},
    pages = {234501},
    year = {2024},
    month = {06},
}

\end{document}